\documentclass{aa}  

\usepackage{graphicx}
\usepackage{txfonts}
\usepackage{rotating}

\begin{document} 

    \title{Comparative analysis of two episodes of strongly geoeffective CME events in November and December 2023}
 

   \author{M. Temmer\inst{1}, M. Dumbovi\'{c}\inst{2}, K. Martini\'{c}\inst{2}, G.M. Cappello\inst{1}, A.K. Remeshan\inst{2}, \\ F. Matkovi\'{c}\inst{2}, D. Milo\v{s}i\'{c}\inst{1}, F. Koller\inst{1,3}, J. \v{C}alogovi\'{c}\inst{2}, R. Susino\inst{4}, M. Romoli\inst{5}}

   \institute{Institute of Physics, University of Graz, A-8010 Graz, Austria\\
              \email{manuela.temmer@uni-graz.at}
         \and
             Hvar Observatory, Faculty of Geodesy, University of Zagreb, Croatia
             \and Department of Physics and Astronomy, Queen Mary University of London, Mile End Road, London E1 4NS, UK
               \and INAF – Osservatorio Astrofisico di Torino
Via Osservatorio 20, 10025 Pino Torinese (TO), Italy
\and Dip. di Fisica e Astronomia, Università di Firenze Via Sansone 1, 50019, Sesto Fiorentino (FI), Italy           }

   \date{Received June 1, 2024; accepted January, 2025}

 
  \abstract
   {In autumn 2023 series of close-in-time eruptive events were observed remotely and measured in-situ. For that period, we study a set of analogous events on the Sun, where several coronal mass ejections (CMEs) were launched partly from the same (active) regions close to a coronal hole. The two episodes of events are separated by a full solar rotation covering the period October 31 -- November 3 and November 27-- 28, 2023. 
   }
   {Both episodes of eruptive events are related to strong geomagnetic storms occurring on November 4--5 and December 1--2, 2023. We point out the complexity for each set of events and aim to understand how the global magnetic field configuration, solar wind conditions, and interaction between the structures relate to these geomagnetic effects. }
   {We use the graduated cylindrical shell (GCS) 3D reconstruction method for deriving the direction of motion and speed of each CME. The GCS results serve as input for the drag-based model with enhanced latitudinal information (3D DBM), facilitating the assessment of its connection to in-situ measurements. This approach significantly aids in the integrated interpretation of in-situ signatures and solar surface structures.} 
   {The first episode caused visible Stable Auroral Red (SAR) arcs, with a Dst index that dropped in three steps down to $-$163\,nT on November 5, 2023. Close in time two CME-related shocks arrived, separated by a sector boundary crossing (SBC), and followed by a short-duration flux rope-like structure. For the second episode, auroral lights were observed related to a two-step drop in the Dst index down to $-$108\,nT on December 1, 2023. A shock from a CME within the magnetic structure of another CME ahead was identified, again combined with a SBC. Additionally, a clear flux rope structure from the shock producing CME is detected. In both events, we observed distinct short-term variations in the magnetic field (``ripples'') together with fluctuations in density and temperature that followed the SBC.}
   {The study presents a comparative analysis of two episodes of multiple eruptive events in November and December 2023. Besides interacting CME structures, we highlight modulation effects in the geomagnetic impact due to magnetic structures which are related to the SBC. These most likely contributed to the stronger geomagnetic impact and production of SAR arcs for the November 4--5, 2023 event. At the Sun, we found the orientation of the heliospheric current sheet to be highly tilted, which might have caused additional effects due to the CMEs interacting with it.}


   \maketitle
%

\section{Introduction}
 
Based on results from cycle 23, the rates of intense geomagnetic storms during maximum/declining solar cycle phases are found to be almost three/two times as high as during the rising phase of a solar cycle \citep[][]{Echer2008}. This is mainly due to numerous eruptive flares occurring close in time and space, resulting in multiple coronal mass ejections (CMEs) that jointly evolve and may interact in interplanetary space \citep[see reviews by e.g.,][]{Lugaz2017,manchester2017,Zhang2021}. Increased number of solar eruptions create extraordinary conditions in near-Earth space, commonly known under the term ``Space Weather'' \citep[see e.g.,][]{Cliver2004,Bothmer07}. Severe impacts on near-Earth space are usually related to the enhanced strength and complexity of the magnetic field in the impacting solar events \citep[see e.g., the ``panoramic'' paper by][presenting in a combined overview our gained knowledge of space physics and space weather over the past 65 years]{tsurutani23}. 

The complexity of the interplanetary magnetic field sweeping over Earth typically increases due to the interaction between multiple large-scale magnetic structures. These cover not only CMEs, but also co-rotating interaction regions (CIRs), formed by high-speed solar wind streams originating from coronal holes (CHs), or the heliospheric current sheet (HCS). It is broadly accepted, when all these components evolving together as compound interplanetary magnetic structures, they cause much stronger geomagnetic effects compared to their isolated appearance \citep[e.g.,][]{Echer2004, Dumbovic15}. It should be noted that the occurrence of successive CMEs or merged interaction regions not always cause stronger geomagnetic effects, especially when related to northward B$_{z}$ components \citep[see][]{burlaga2002,burlaga2003}. Propagating shocks caused by energetic CMEs, when penetrating through a CME ahead are prone to enhance the magnetic field complexity \citep{Lugaz2016}. In-depth case studies investigate the process of CME-CME interaction \citep[e.g.,][]{gopalswamy01,Temmer12,Scolini2020b}, or CME-CIR interaction \citep[e.g.,][]{fenrich98,Heinemann2019,Winslow2021,geyer23}. However, the details of the interaction processes between these different structures are still not well understood. This complicates the interpretation of in-situ measurements and the development of reliable models for forecasting geomagnetic effects.

Compared to 20 years ago, during the peak of solar cycle 23, the coverage (multiple viewpoints, varying distances including heliospheric imagers) and quality (temporal and spatial resolution) of solar data have improved substantially. These advancements have been accompanied by significant progress in our investigation techniques and modeling approaches. Solar cycle 24 was notably weak \citep[e.g.,][]{McComas13}, which made it challenging to test our understanding of complex solar activity phenomena. The current solar cycle 25 appears to be more similar to cycle 23 in terms of geomagnetic effects and the complexity of the related phenomena \citep[see e.g.,][]{berdichevsky00,burlaga2002,burlaga2003}. Recent events occurring at the end of October and beginning of November 2023 produced intricate in-situ measured profiles together with strong auroral light activities. In addition to auroral lights, a special type of atmospheric light phenomena, so-called Stable Auroral Red (SAR) arcs \citep{Cole65}, were observed almost globally on November 5, 2023 reaching even southern Europe. SAR arcs are known to be related to extreme conditions in the Earth's ring current system \citep[see review by e.g.,][]{kozyra03}. Besides SAR arcs, on November 5, 2023 also a very rarely observed anisotropic cosmic-ray enhancement (ACRE) was reported by \cite{Gil2024}. A full rotation later, end of November 2023, again a series of CMEs was launched heading towards Earth. These events resulted in strong geomagnetic effects, but to a comparatively weaker Disturbance Storm Time index (Dst of $-$108\,nT for December 1, 2023 versus $-$163\,nT for November 5, 2023) and no SAR arcs.

In the present study we investigate these two episodes of CME events, to understand their similarities as well as differences. The two episodes are united by the involvement of similar (active) regions and the same CH, but separated in time by a complete solar rotation. We perform a thorough investigation of the solar source regions, CME development and potential interaction with the nearby high-speed stream emanating from the CH. This is complemented by modeling efforts to better match and interpret in-situ measurements, focusing on the arrival time and impact of each event. For modeling the interplanetary propagation behavior of the CMEs we apply the drag-based-model \citep[DBM;][]{Vrsnak2013} in an enhanced 3D version, i.e., covering latitudinal information (3D DBM; Dumbovic et al., 2024, to be submitted). We will highlight the importance of HCS orientation and crossings, which enhance the complexity of solar activity phenomena and their impact on planets \citep[see also COSPAR Space Weather Roadmap update by][]{Temmer2023}. 

The paper is structured as follows. In Section~\ref{solar} we investigate the solar perspective and use that results to run the 3D DBM propagation model. With this we assess the impact likeliness and arrival time of each CME at Earth for better linking and interpretation of the in-situ signatures. In Section~\ref{earth} we analyze the in-situ measurements for each rotation and relate to the geomagnetic impact. In Section~\ref{conclusion} we present a discussion and our conclusions. Appendices~\ref{appendix} and \ref{appendix2} cover complementary material for the analysis and methodology, respectively.

\section{The solar perspective}\label{solar}

\subsection{Data and Methods}\label{CHIP}
   \begin{figure*}
   \centering
   \includegraphics[width= 0.95\textwidth]{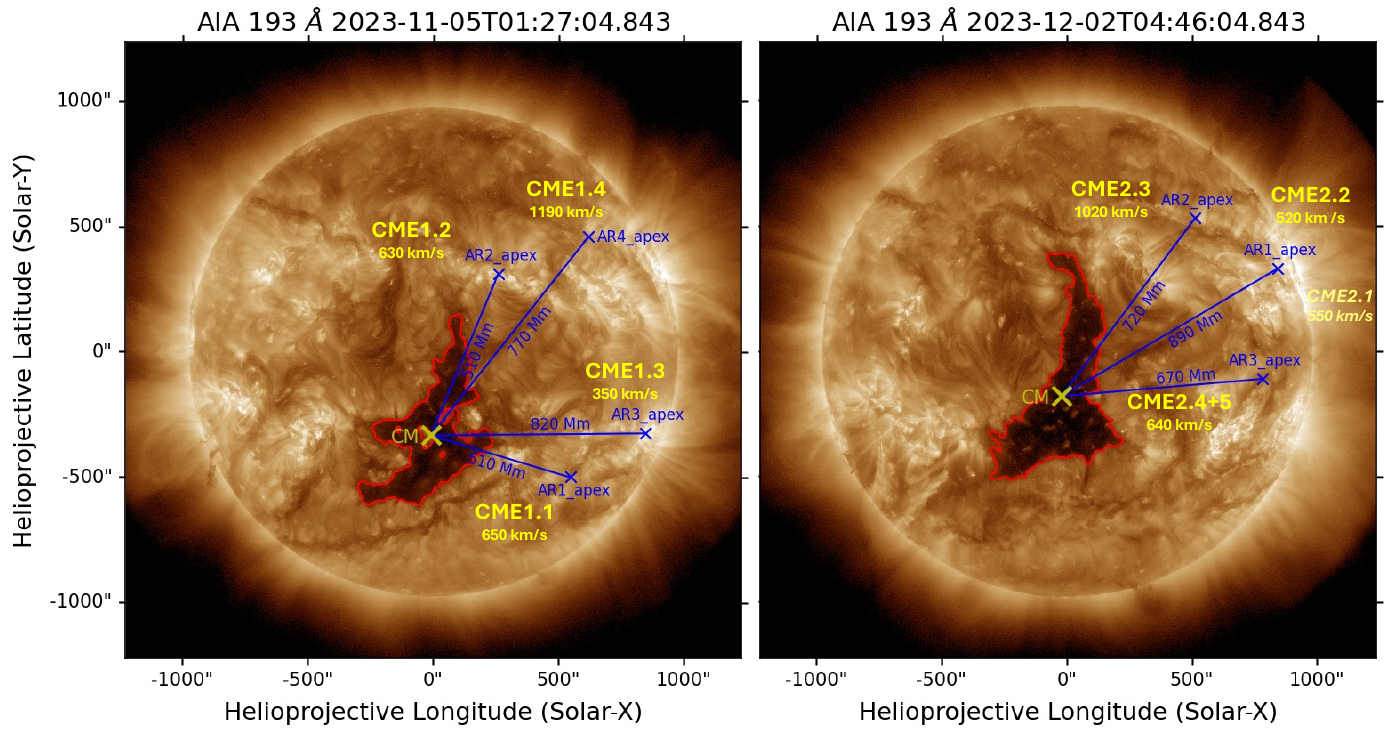}
      \caption{General overview of the CME occurrence and CHs for the two episodes, rotation~\#1 (left panel) and rotation~\#2 (right panel), overlaid on AIA 193\AA~images. Given are CH locations with their center of mass (CM) and apex distances to the active regions (AR) from which the CMEs were launched (more details are given in Section~\ref{CHIP}). For each CME the 3D speed information is given as derived from a linear fit to sequential GCS reconstructions. For rotation~\#1 we observe four CMEs (CME1.1--1.4) and for rotation~\#2 five CMEs (CME2.1--2.4+5). We note that for CME 2.1 the corresponding active region is already behind the western limb and therefore no AR distance is given. Image dates were chosen such that the CH is centrally located for better visibility. 
}
         \label{overview}
   \end{figure*}
%

We investigate two episodes of multiple eruptions from the Sun covering the time range October 31--November 3, 2023 (rotation~\#1) within Carrington rotation 2277 and November 27--28, 2023 (rotation~\#2) within Carrington rotation 2278. Separated by a full solar rotation, the two episodes of events show many similarities, as partly the same (active) regions are involved, which are located primarily West from the same CH (see Figure~\ref{overview} with more details given in Sections~\ref{solar-rot1} and~\ref{solar-rot2}). 

To study the solar surface structures of the erupting events, related filaments and flares as well as their magnetic properties, we use data from the Solar Dynamics Observatory \citep[SDO;][]{Pesnell2012}. These cover EUV images from the Atmospheric Imaging Assembly \citep[AIA;][]{Lemen2012} together with photospheric magnetic field information from the Helioseismic and Magnetic Imager \citep[][]{Scherrer2012}. Figure~\ref{overview} provides an overview of the eruption sites of the CMEs relative to the CH for rotation~\#1 and \#2, superimposed on an AIA 193\AA~image. For each evolving CME we apply on stereoscopic white-light coronagraph data the graduated cylindrical shell model \citep[GCS;][]{Thernisien2006}. GCS simulates the flux rope structure that is often associated with CMEs. Fitting the model to coronagraph images from at least two different viewpoints enables to obtain the 3D CME geometry and propagation direction. Applying that procedure over several time steps allows to derive the 3D speed. Error estimates for the height-time measurements from GCS reconstructions and derived speeds are of the order of 5\% as given by \cite{Verbeke2023}. We use time-series of white-light data from the Large Angle Spectroscopic Coronagraph aboard the Solar and Heliospheric Observatory \citep[SOHO/LASCO C2 and C3;][]{Brueckner1995} and the COR2 coronagraph from the Sun Earth Connection Coronal and Heliospheric Investigation \citep[SECCHI;][]{Howard2008} aboard the Solar Terrestrial Relation Observatory  \citep[STEREO-A;][]{kaiser08}. During that period we had only a small separation between STEREO-A and SOHO (5--7$^\circ$). For better constraining the GCS forward model, we added Metis coronagraph data \citep[visible light VD channel at 610~nm;][]{Antonucci2020, Fineschi2020, DeLeo2023} aboard Solar Orbiter \citep{Mueller2020}. Solar Orbiter is located East of Earth with a separation of $\sim$20$^\circ$ for rotation~\#1 events and $\sim$10$^\circ$ for rotation~\#2 events. 

CHs are the main sources of fast solar wind emanating along the open magnetic field within the CH and interacting with the slow solar wind ahead, forming CIRs \citep[see e.g., review by][]{Cranmer2017}. The CH under study, which is the same for all the events, is investigated using CATCH \citep{heinemann2019_CATCH}. Results from CATCH and GCS are used to calculate the so-called coronal hole influence parameter (CHIP). CHIP assesses the influence of the CH on each CME. The CHIP is calculated from the line-of-sight magnetic field strength and area of the CH in relation to the CME's source region distance \citep[see][and references therein]{Gopal2009}. The CHIP value (in Gauss) is given for the distance between the CH center of mass and the projection of the CME apex on the solar disc. The derived distances marked by ``AR$\textunderscore$apex'' are given in Figure~\ref{overview}. The CHIP reveals the impact of the CH on the behavior of CME propagation, particularly its potential to divert the CME magnetic structure from its radial trajectory. 

Utilizing the results from the GCS reconstruction, we can generate a projected view of each CME using a straightforward cone geometry, considering both equatorial and meridional planes (equations outlined in Dumbovic et al., 2024, to be submitted). The CME cone projections enable us to more easily (or better) connect remote sensing imagery with in-situ data. (This is obviously possible to do with the projection or with other tools). This allows us to determine which CME, and more specifically which region of the CME (apex, flank), is most likely associated with the in-situ measurements. Covering the GCS error estimates, we calculate the best- and worst-case impact scenarios representing conditions that are either more favorable for a hit or more likely to result in a miss. 

Table~\ref{summary_table} gives a summary for each CME and its source region characteristics (block 1), CATCH and CHIP results for the CH (block 2), and 3D DBM hit/miss statistics (block 3). A summary of all derived GCS parameters, their errors and hit/miss likeliness is given in Table~\ref{tiny_table} in Appendix~\ref{appendix}.

\begin{sidewaysfigure*}%
  \includegraphics[width=22cm]{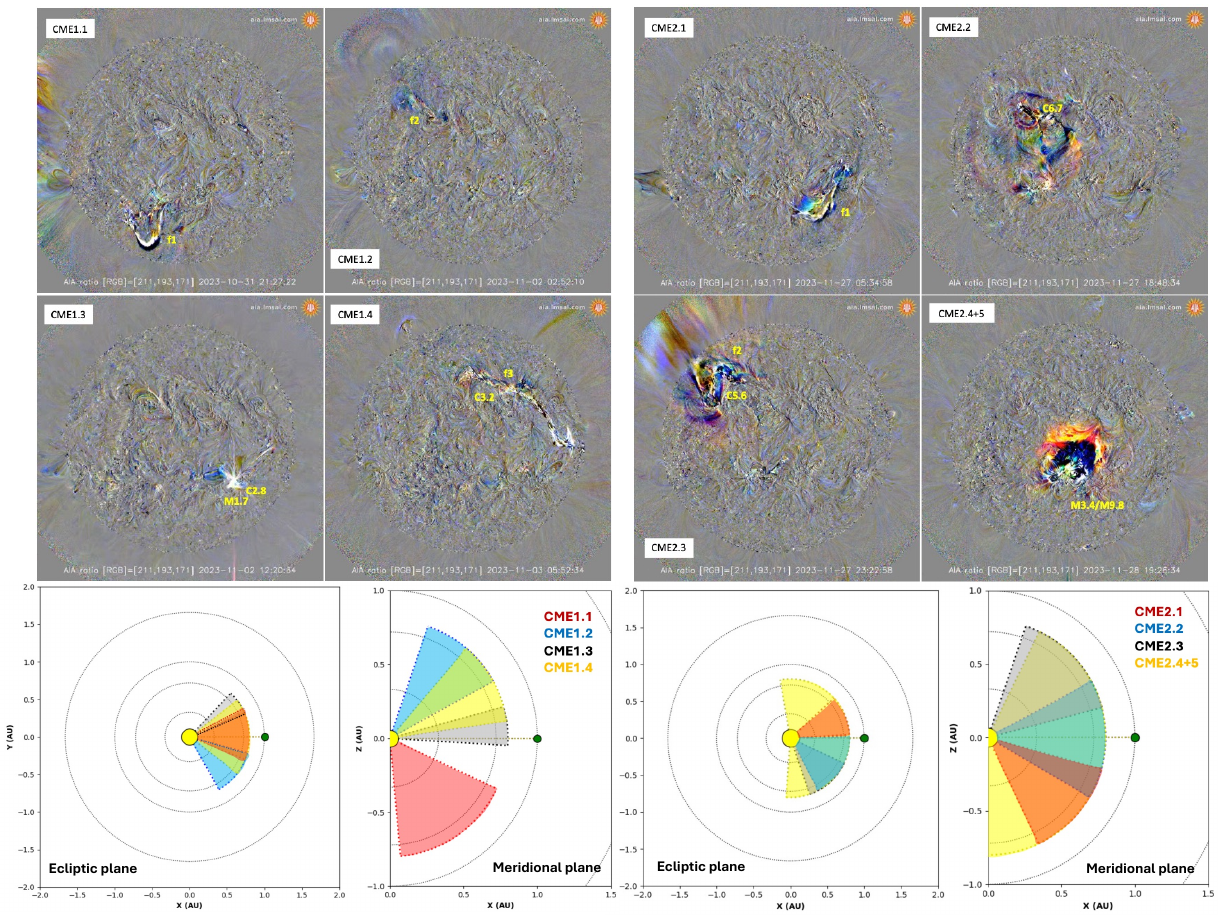}
      \caption{Top panels: SDO/AIA EUV composite ratio images (source: suntoday.lmsal.com) for each eruptive event. Bottom panels: CME cone geometry derived from GCS results (longitude, latitude, tilt) projected on the ecliptic and meridional plane. The events are annotated with the eruption related GOES SXR flare classes. Left panels: rotation~\#1 over the period 2023-10-31 21:27~UT until 2023-11-03 05:52~UT. CME1.1+1.2+1.4 are related to filament eruptions f1+f2+f3, respectively. Right panels: rotation~\#2 over the period 2023-11-27 18:48~UT until 2023-11-28 19:50~UT. CME2.1+2.4 are related to filament eruptions f1+f2, respectively.}
         \label{over2}       
   \end{sidewaysfigure*}
   

\subsection{Near-Sun results for rotation~\#1}\label{solar-rot1}
\begin{figure*}
   \centering
   \includegraphics[width= 0.9\textwidth]{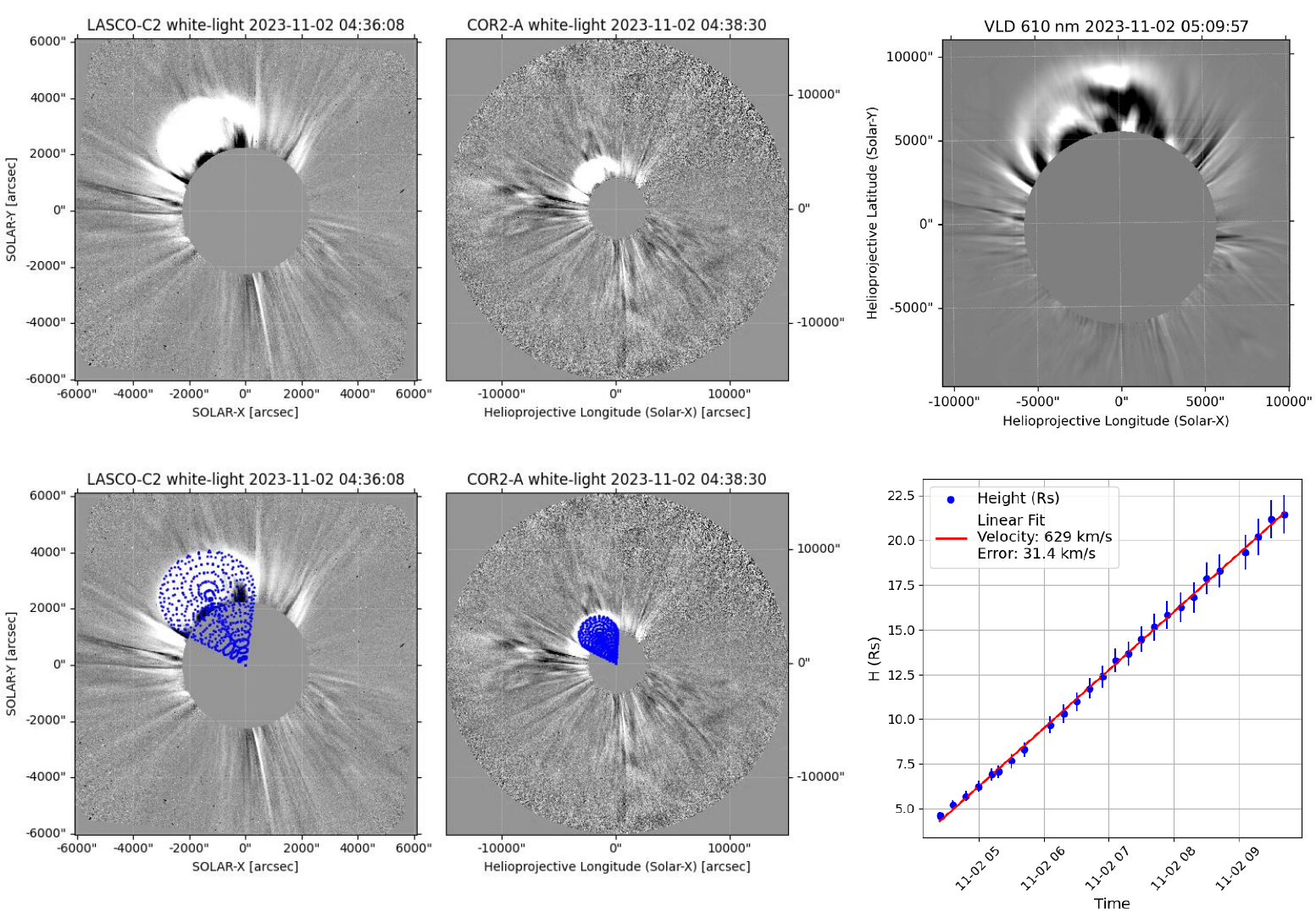}
      \caption{For rotation~\#1 we show on the example of CME1.2 the CME structure from different perspectives (LASCO C2, COR2 and Metis) and GCS reconstruction for the CME at a 3D height of 5.4Rs. Bottom right panel: from different time steps the height-time and speed are derived using a linear fit (given in the legend). 
}
         \label{CME1.2}
   \end{figure*}

During the first rotation, we focus on the occurrence of four prominent CMEs, termed CME1.1, 1.2, 1.3, and 1.4, respectively. For rotation~\#1, the top left panels of Figure~\ref{over2} provide an overview of the eruption sites. The bottom left panels show the projected cone geometries for each CME as derived from the GCS results. In addition to LASCO C2 and COR2 data, Metis observations were available for CME1.1, CME1.2, and CME1.4. We present in Figure~\ref{CME1.2} the GCS reconstruction and kinematics on the example of CME1.2.  All other GCS reconstruction plots from rotation~\#1, together with the height-time plots, linear fits, and speed derivation, are given in Figures~\ref{CME1.1} to \ref{CME1.4} in Appendix~\ref{appendix}. 

CME1.1 started from a filament eruption at around 2023-10-31 20:00~UT, located in the southern hemisphere spanning heliographic coordinates S18--S40/E00--E40. The eruption site is not related to an active region. The CME is observed in white-light image data with a first appearance in LASCO C2 at 2023-10-31 22:12~UT at the position angle (PA) 160--200°. Tracking the CME over time and performing GCS reconstruction for each time stamp we derive a speed of  $\sim$650$\pm$30 km/s. From the orientation of magnetic tongues, filament barbs, and post-eruptive arcades (PEA) it can be related to a right-handed chirality \citep[see e.g.,][]{Palmerio2018}. The projected view of the CME in simple cone geometry in the meridional cut reveals that the CME primarily moves to the South. From the best-hit scenario CME1.1 would be a only glancing blow.

CME1.2 is related to a filament eruption at 2023-11-02 ca.\,03:00~UT, located in the northern hemisphere at the heliographic coordinates N25/E25. The eruption site is not related to an active region. The CME is observed in white-light image data with a first appearance in LASCO C2 at 2023-11-02 03:36~UT at PA 0--70°. From GCS reconstructions we calculate a speed of $\sim$630$\pm$30 km/s. The chirality is derived as left-handed. The projected cone geometry from GCS shows that the CME moves far away from Sun-Earth line directed to North-East. In the best-hit scenario CME1.2 would miss Earth.

CME1.3 is not related to a filament eruption but to an active region (AR 13474) that produced two flares with about 8 hours time difference: an M1.7 (GOES SXR-class) flare peaking at 2023-11-02 12:22~UT and a C2.8 flare with its maximum at 20:18~UT. The flare location is given at the heliographic coordinates S18/W30-34. In LASCO C2 the CME front appeared first 2023-11-02 23:12~UT at PA 260--280°. Though the CME itself is weak with a speed of  $\sim$350$\pm$20 km/s as derived from GCS reconstruction, it seems to have destabilized the filament related to CME1.4 erupting the following day. The chirality for this complex two-flare source region could not be unambiguously derived, as neither magnetic tongues, sigmoids, post-eruptive flare ribbons, nor filament barbs could be identified. The projected cone geometry from GCS shows that the CME propagates in the western direction, away from the Sun-Earth line. In the best-hit scenario CME1.3 would be a flank hit.

CME1.4 is produced due to a large filament eruption starting to rise at 2023-11-03 06:00~UT in the northwestern hemisphere spanning a heliographic range of N00-N30/W00-W60. The source region is related to AR 13473. The filament rise is preceded by a C3.2 flare starting at 2023-11-03 04:40~UT at N31W26. The first LASCO C2 image showed the CME front at 2023-11-03 05:48~UT and it continued to evolve as partial halo event. From GCS reconstructions we derive a speed of $\sim$1190$\pm$60 km/s. The chirality for the flux rope is derived as left-handed. From the projected GCS cone we see that the CME flux rope part propagates slightly northwards from Earth. Applying nominal GCS parameters and those producing a best-hit, CME1.4 would result in a glancing blow or even full hit for Earth.   

The CH polarity is derived to be negative (signed/unsigned mean magnetic field strength of $-$5.2\,G/12.2\,G). The size in area of the CH (8.5$\times$10$^{10}$ km$^2$) would be related to a high-speed stream with a speed at 1AU in the order of $\sim$750\,km/s \citep[see][]{vrsnak2007}. Using these results, the calculated CHIP parameter for each CME reveals values well below 2.6\,G, which was found as a lower limit corresponding to the deflecting effect by a CH \citep{Makela2013}. 


To summarize, speeds for CME 1.1-1.3 are found to be in the slow to moderate range with values at 21.5\,Rs in the order of 350--650 km/s. CME1.4 is the most energetic event, revealing  $\sim$1190\,km/s at 21.5\,Rs, and is related to a huge filament eruption. Due to the low CHIP values found, we do not expect any of the CMEs to get significantly deflected from their direction as derived from low coronal signatures. Interpreting the source surface information, 3D geometries of the CMEs, and reconstructed speeds, we would conclude that only CME1.4 is a potential candidate to partially hit Earth, with CME1.3 as most likely additional candidate. Being a miss even in the best-hit scenario we confidently exclude CME1.2 as a candidate for appearing in the in-situ signatures and would categorize CME1.1 as unlikely to cause significant in-situ signatures.


\subsection{Near-Sun results for rotation~\#2}\label{solar-rot2}
\begin{figure*}
   \centering
   \includegraphics[width= 0.9\textwidth]{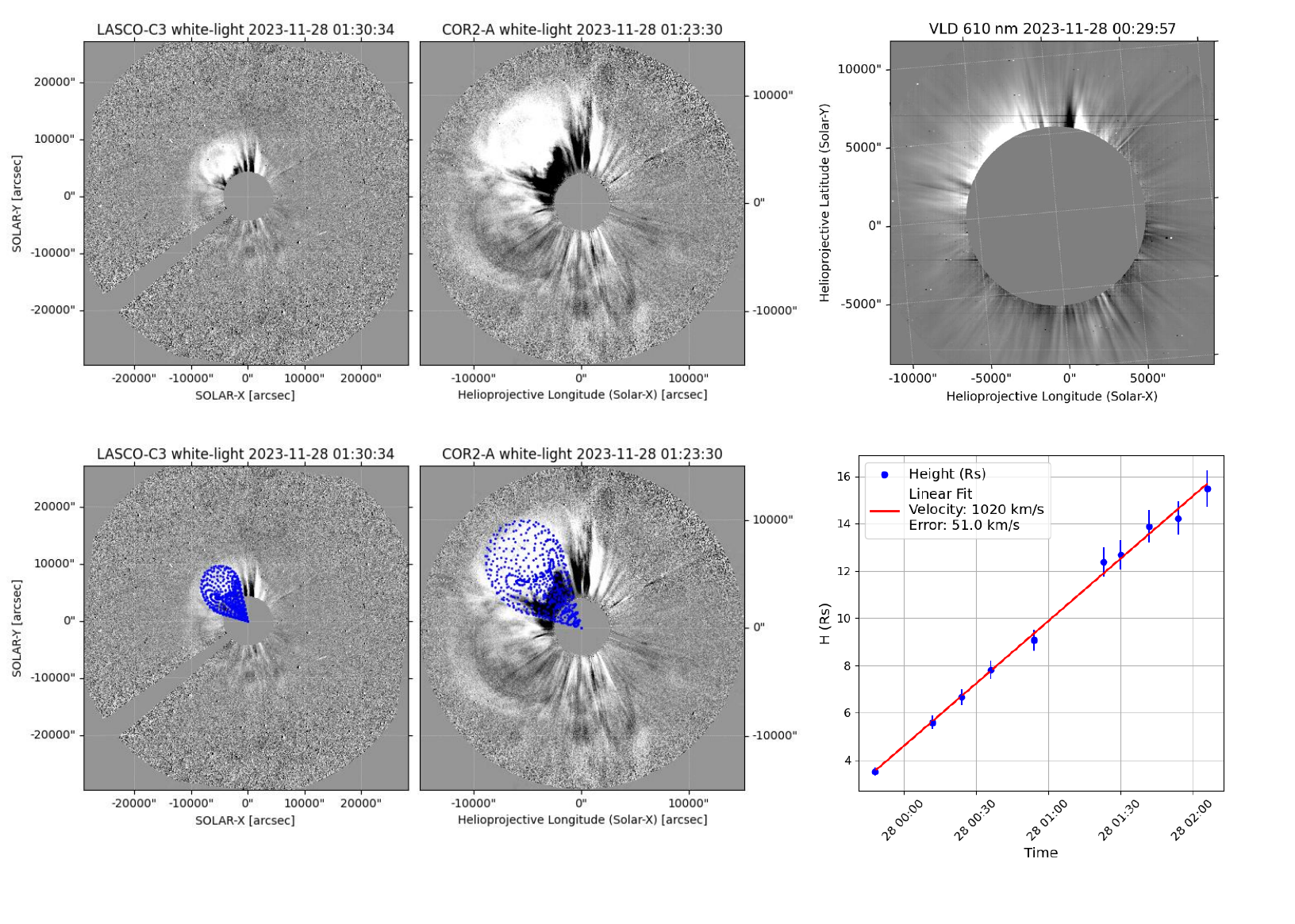}
      \caption{For rotation~\#2 we show on the example of CME2.3 the CME structure from different perspectives (LASCO C2, COR2 and Metis) and GCS reconstruction for the CME at a 3D height of 13.0Rs. Bottom right panel: from different time steps the height-time and speed are derived using a linear fit (given in the legend).
}
         \label{CME2.3}
   \end{figure*}

During the second rotation, we focus on the occurrence of five CMEs. For rotation~\#2, the top right panels of Figure~\ref{over2} provide an overview of the eruption sites. The bottom right panels show the projected cone geometries for each CME as derived from the GCS results. In addition to LASCO C2 and COR2 data, Metis observations were available for CME2.1, CME2.3, and CME2.4+5. We present in Figure~\ref{CME2.3} the GCS reconstruction and kinematics on the example of CME2.3. The GCS reconstruction plots for all other CMEs from rotation~\#2, together with the height-time plots and linear fits, as well as speed derivation, are given in Figures~\ref{CME2.1} to \ref{CME2.4+5} in Appendix~\ref{appendix}.  

CME2.1 is related to a filament eruption (close to AR 13499) starting at 2023-11-27 ca.\,05:00~UT from heliographic coordinates around W30/S30. It appears in LASCO C2 images at 2023-11-27 06:48~UT at PA 150--250°. Performing GCS reconstructions we derive a speed of $\sim$550$\pm$30 km/s. The chirality is derived from the orientation of the PEA and underlying magnetic field as right-handed. The projected cone geometry from GCS shows that the CME moves far away from Sun-Earth line directed to the South but might affect subsequent CMEs due to its large width in longitude. From the best-hit scenario, CME2.1 would be a flank hit.

CME2.2 is related to a C6.7 flare starting 2023-11-27 18:52~UT at N18E09 (east from AR 13503) with further activity signatures connecting to the South (AR 13500), presumably via some global magnetic field that might got disturbed as CME2.2 erupted. It appears in the LASCO coronagraphs as partial halo revealing its front in the C2 field-of-view first time at 2023-11-27 20:12~UT. Tracking the CME over time and performing GCS reconstructions we derive a speed of $\sim$520$\pm$30 km/s. The chirality is derived as left-handed. The projected cone geometry from GCS would suggest a direct impact on Earth, with the hit occurring relatively close to the CME apex. From the best-hit scenario, CME2.2 would be a full apex hit.

CME2.3 is related to a filament eruption and C5.6 flare at 2023-11-27 23:40~UT located at N26E40. The source region is located west from AR 13503. In LASCO C2 it appears first at 2023-11-27 23:48~UT at PA 0--90°. From GCS reconstructions we derive a speed of $\sim$1020$\pm$50 km/s. The chirality is derived as left-handed. The projected cone geometry from GCS shows that the CME moves far away from Sun-Earth line directed to North-East. For the best-hit scenario, CME2.3 would miss the Earth.

CME2.4+5 is related to an extended double eruptive event on 2023-11-28 with two strong flares (M3.4+M9.8) from a source region located at heliographic coordinates S16W00 (AR 13500). The observed signatures of a coronal wave indicate a strong lateral expansion of the CME \citep[see e.g., review by][]{Patsourakos2012}. The M3.4 flare from 2023-11-28 peaked at 19:32~UT immediately followed by a M9.8 flare with a maximum at 19:50~UT. The CME is reported in LASCO as a full halo and the double-eruption shows fronts close in time at 2023-11-28 20:24~UT and 20:48~UT. From the short time difference we may safely presume that the CMEs started to interact close to the Sun and therefore will be treated as a single eruption. Tracking the outermost CME front we derive from GCS reconstructions a speed of $\sim$640$\pm$30 km/s. The chirality for the eruptions from that source region is derived as right-handed. The projected cone geometry from GCS reveals a hit due to its huge extent. Even for the worst-hit scenario CME2.4+5 would still be classified as a hit.

The CH polarity is found to have decreased in its magnetic field strength (signed/unsigned mean magnetic field strength of $-$3.5\,G/10.9\,G). On the other hand, the area is found to have increased to 10.1$\times$10$^{10}$ km$^2$. This non-correlation between area and magnetic field strength supports findings by \cite{Heinemann2020} who concluded that the magnetic flux within a CH is not the main cause for its evolution. Again we derive for each of the CME events very low CHIP values, indicating low or no influence of the CH on any of the CME trajectories.

In summary, CME 2.1, 2.2, and 2.4+5 (double eruptive event) show rather moderate speeds with 520--640 km/s at the height of 21.5\,Rs while CME2.3 is derived with $\sim$1020 km/s. Interpreting the source surface information, 3D geometries of the CMEs, and reconstructed speeds, we would conclude that CME2.2 and CME2.4+5 most likely hit Earth. Since CME2.3 is a miss even under the best-hit scenario, we confidently rule it out as a candidate for being detected as in-situ signature. CME2.1 is identified as unlikely to cause in-situ signatures.

\section{Interplanetary space perspective}\label{earth}

\subsection{Data and Methods}\label{subsection:earth}


The in-situ measurements are taken from the OMNI\_HRO\_1MIN dataset, which is provided by the Coordinated Data Analysis Web\footnote{\url{ https://omniweb.gsfc.nasa.gov}}. The data cover 1-minute resolution time series of solar wind plasma parameters and interplanetary magnetic field components in the geocentric solar ecliptic (GSE) coordinate system, shifted to the nose of the Earth's bow shock \citep[][]{omni}. From the plasma and magnetic field components we calculated the plasma-$\beta$ value.

To study and assess the geo-effectiveness, we use the Dst and Hpo indices. The Dst index measures the average deviation in the horizontal component of the Earth's magnetic field as well as the plasma energy content of the inner magnetosphere at mid-latitude stations around the globe \cite[see e.g.,][]{Turner2001}. For the analysis we used the provisional Dst index\footnote{\url{https://wdc.kugi.kyoto-u.ac.jp/dst_provisional/}; the final Dst index was not available at the time of writing.}  with a 1-hour time-resolution. The geomagnetic Hpo index describes in a similar way as the Kp index the disturbance levels in the two horizontal magnetic field components. In comparison to Kp with 3-hours time resolution scaling from 0 to 9, the Hpo index is derived with a higher time resolution of 30 and 60 minutes, respectively, and its scale is open ended. For our studies we use Hp30\footnote{\url{https://kp.gfz-potsdam.de/en/hp30-hp60}} covering a temporal resolution of 30 minutes \citep{Yamazaki22}.  

To establish the connection between results derived from remote sensing image data and in-situ measurements, we utilize the analytical drag-based model (DBM). Simulations were conducted for CME1.4, CME2.2, and CME2.4+5, which were identified as potential candidates for exhibiting in-situ signatures. The results from the GCS model (3D geometry, speed, and direction of motion) are used as input parameters for deriving the arrival time and speed of the flux rope (i.e., magnetic structure) of the CME \citep[][]{Vrsnak2013}. DBM was recently modified for 3D CME geometry (3D DBM; Dumbovic et al., 2024; to be submitted). DBM usually uses GCS results to calculate the extent of the CME in the solar equatorial plane and runs the CME leading edge as a non-self-similarly expanding 2D cone \citep{Dumbovic2021_DBM}. 3D DBM takes into account the heliographic latitude or the extent of the CME in the solar meridional plane, applying the geometry of the non-self-similarly expanding 2D cone in both the solar equatorial and the meridional plane (cf.\ bottom panels of Figure~\ref{over2}). Moreover,  3D DBM can be used to model CME-CME interaction in a very simple manner \citep[shown by e.g.,][using DBM]{Guo2018,Dumbovic2019} as outlined in Appendix~\ref{appendix2}. We acknowledge that this is a simplified approach, however, it serves the purpose of linking CMEs to their in-situ measurements. For a more comprehensive and physically detailed analysis of CME-CME interactions, we refer to advanced MHD models by e.g., \cite{Lugaz2005} or \cite{Scolini2020b}.

Table~\ref{summary_table} gives a summary of the 3D DBM results for shock and flux rope arrival time (block 3) as well as in-situ measured times for comparison (block 4). Related GCS input parameters to run the 3D DBM are given in Table~\ref{tiny_table} in Appendix~\ref{appendix}.

\subsection{In-situ results for rotation~\#1}\label{earth-rot1}

   \begin{sidewaysfigure*}
   \centering
   \includegraphics[width=20cm]{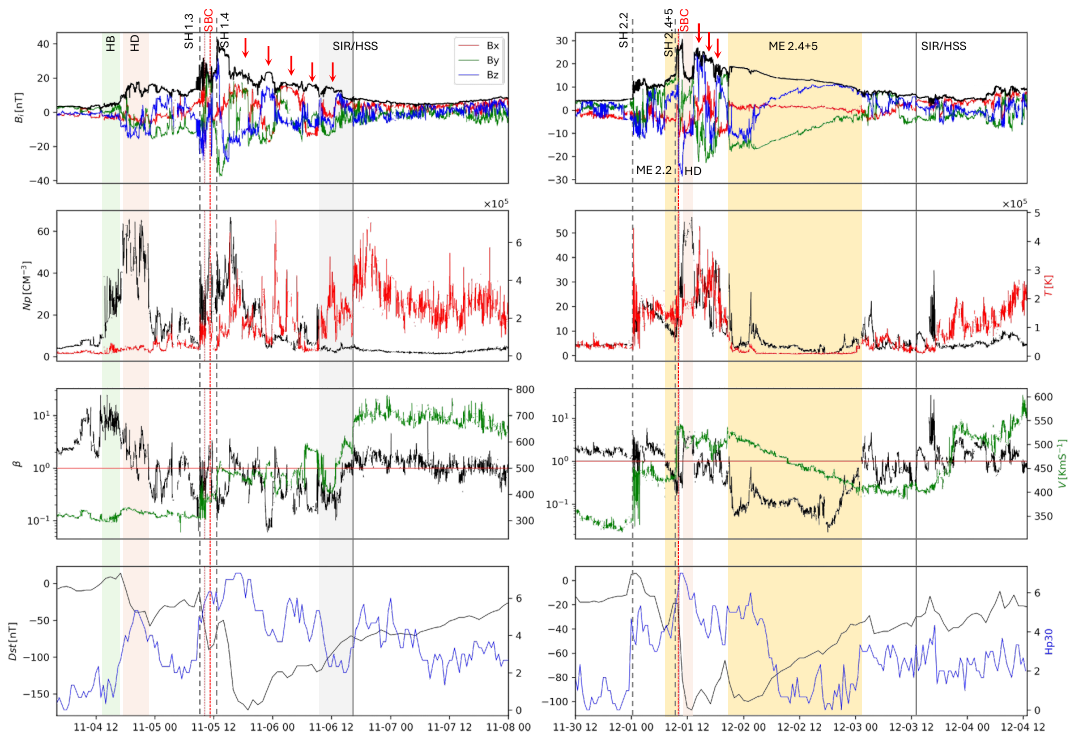}
      \caption{OMNI in-situ measurements and geomagnetic indices for each rotation (the x-axis shows time in UT). From top to bottom: magnetic field/vector components, density+temperature, plasma-$\beta$+speed, Dst+Hp30. We mark a high plasma-$\beta$ structure heliospheric plasma sheet (HB), high-density (HD) structures, the shocks (SH; dashed vertical lines) from various CMEs, the sector boundary crossing (SBC; dotted red vertical line), magnetic ejecta (ME) related to the different CMEs, and the start of the SIR/HSS (gray shaded area marks the ''interaction´´ zone between the rear-part of the CME and the starting of the SIR). Red arrows indicate the observed ripples (short-term variations in the total magnetic field associated with strong fluctuations in the temperature and density profiles). A zoom-in for the regions around the SBC is given in Figure~\ref{zoom-insitu} in Appendix~\ref{appendix}.
}
         \label{insitu}       
   \end{sidewaysfigure*}
 
Results from the projected cone geometries show that CME1.1 and 1.2 miss Earth and CME1.3 most likely misses Earth. Only CME1.4 is derived to partly hit Earth. Based on these results, we would therefore not expect clear signatures of a magnetic ejecta (ME) in the in-situ measurements from any of the CMEs. For CME1.4 we calculated with 3D DBM the arrival time with November 5, 2023 at 23:00~UT, with a tentative shock-sheath arrival 10 hours earlier, i.e., 13:00~UT. We note that this tentative shock arrival time is estimated based on the typical shock/sheath duration from a statistical approach as given by \cite{Russell2002}.




The left panel of Figure~\ref{insitu} shows an overview of Earth-related measurements for rotation~\#1. In-situ measurements reveal two shocks very close in time at November 5, 2023 at 09:09~UT and 12:38~UT. In between these two shocks a sector boundary crossing (SBC) is identified (a list of SBCs is maintained by Leif Svalgaard\footnote{\url{https://svalgaard.leif.org/sblist.txt}}). SBCs mark the polarity transition across the wavy HCS, which is embedded in a much larger plasma sheet structure \citep{Winterhalter1994}. The HCS signatures (plasma density, total magnetic field) associated with the SBC are not fully clear and could also be interpreted as deflection patterns in the interplanetary magnetic field caused by the CME shocks, likely corresponding to the flank crossings of CME1.3 and CME1.4. The second shock is followed by a prominent short-duration magnetic field structure (12:38--15:09~UT) of low plasma-$\beta$ (cf.\,Figure~\ref{zoom-insitu} for a zoom-in version of Figure~\ref{insitu}). Afterwards we observe some distinct magnetic field variations that we refer to as ``ripples'' (see more details below). November 4, 2023, a day before the first shock-arrival, a high plasma-$\beta$ region is measured. That high plasma-$\beta$ structure is followed by a huge increase in density up to $\sim$60\,N$_p$ per cm$^{3}$ and a negative B$_z$ component of about $-$18\,nT. The region hints towards the extended heliospheric plasma sheet (HPS; streamer belt plasma) surrounding the current sheet \citep{Burlaga1990, Winterhalter1994}. Though the HPS is typically not related to strong out-of-the-ecliptic field, the closeness to the CME shock might have compressed the HPS. A similar scenario was observed and modeled by \cite{Wu17}. That specific solar wind structure is related to the first drop in Dst down to $-$54~nT at November 4, 2023 around 24:00~UT. 

The arrival of the two CME-related shocks (most likely related to CME1.3 and CME1.4, respectively) cause a slight increase in Dst again, each followed by a second and third drop down to $-$83\,nT (November 5, 2023 at 11:00~UT) and $-$163\,nT (November 5, 2023 at 19:00~UT), respectively. The second drop in Dst is related to strong fluctuations in between the shock-sheath part of CME1.3 and SBC region with a minimum B$_z$ of $-$24\,nT. The third strong Dst drop starts with the arrival of the short-duration magnetic field structure having a minimum B$_z$ of $-$28\,nT and continues to drop further during the ripples. Compared to that, the Hp30 index reached on November 5, 2023 a peak value of 5.33 at 20:00~UT and further increased to 7.00 at 14:30~UT, with peak values of 7.33 between 16:30 and 17:30~UT (low-latitude SAR arcs were reported around 17:30~UT). To conclude, we can attribute the three-step drop in Dst to the occurrence of a high-density and rather strong negative B$_z$ streamer belt plasma related to the HPS, shock-sheath magnetic field fluctuations and specific magnetic field configurations influenced by the sector boundary. A three-step decrease in the Dst is also reported for the complex events in September 2017 by \cite{Hajra2020}. 

As expected the SIR arrives later on November 6, 2023 at 09:30~UT (marked by a gray area in Figure~\ref{insitu}) with a change in the azimuthal flow angle (not shown), followed by the HSS around 16:30~UT. Interestingly, most of the observed ripples are located in the region before the SIR/HSS and after the shock arrival from CME1.4 that follows the SBC. These kind of structured signatures reveal short-term variations in the mesoscale range of a few hours in the total magnetic field with either correlated or anti-correlated profiles among the vector components. The small-scale variations are separated by abrupt changes in the magnetic field orientation. The structures are related to strong fluctuations in the temperature and density profiles. We marked the ripples with red arrows in Figure~\ref{insitu}.


\subsection{In-situ results for rotation~\#2}\label{earth-rot2}

CME cone projections would predict for the magnetic structures misses for CME2.1 and 2.3, a clear apex hit for CME2.2 and a tentative hit for CME2.4+5. To give a rough estimate if an interaction between CME2.2 and CME2.4+5 is likely, we performed a two-step 3D DBM run. We use as input the GCS results within the error estimates, and for the following CME we lower the $\gamma$ value almost by a factor of two and enhance the solar wind speed. From that we obtain a hint towards a likely interaction between CME2.2 and CME2.4+5. These may arrive at Earth as a combined entity on December 2, 2023, at approximately 00:44~UT, preceded by a tentative shock on December 1, 2023, at around 14:44~UT. Consequently, as CME2.2 and CME2.4+5 likely arrive in close temporal proximity we would expect in the in-situ measurements ME signatures from both CMEs reflecting a complex structure.  



The right panel of Figure~\ref{insitu} shows an overview of Earth-related measurements for rotation~\#2. The shock arrival of CME2.2 is detected in OMNI December 1, 2023 at 00:22~UT and for CME2.4+5 at 09:38~UT. The ME of CME2.2 is identified at 07:28~UT (with a duration until 10:07~UT) and for CME2.4+5 at $\sim$20:45~UT. Comparing 3D DBM results to in-situ measurements we derive that the modeled shock arrival time for both CMEs is delayed by 6--15.5~h. Besides the simplistic interaction scenario simulated with the two-step 3D DBM, this could be due to preconditioning effects \citep[see e.g.,][]{Temmer2017} as well as the following HSS influencing the CME propagation behavior. Only for CME2.4+5 clear signatures of a flux rope are revealed from which the FR chirality can be determined, that matches the one derived from the solar surface structures (right-handed). A SBC is reported December 1, 2023 at 10:11~UT (see also Figure~\ref{zoom-insitu} for a zoom-in version of Figure~\ref{insitu}). That SBC is clearly related to the HCS revealing a spike in density and a drop in the total magnetic field. 

The geomagnetic effects are moderate-to-strong with a two-step drop in Dst of $-$39\,nT December 1, 2023 at 08:00~UT followed by $-$108\,nT at 14:00~UT. The increase in Hp30 and decrease in Dst starts with the arrival of the first shock presumably related to CME2.2. In the sheath of CME2.2 we observe fluctuations in the interplanetary magnetic field  which are related to a minimum B$_z$ of $-$12\,nT. Dst and Hp30 reach a first minimum/maximum as the ME of CME2.2 reaches Earth. The second shock signature presumably related to CME2.4+5 is observed to lie within that magnetic structure, causing a slight increase in Dst again. The magnetic field that follows is very complex and is compressed into the trailing part of CME2.2 and a SBC resulting in a minimum B$_z$ of $-$28\,nT, and followed by a high-density region. This caused the Dst to drop further, down to its minimum value of $-$108\,nT. The incoming ME from CME2.4+5 did not cause further enhancement in Hp30 or Dst. The geomagnetic indices remain rather low/high (Dst/Hp30) until B$_z$ changes sign within the ME December 2, 2023 around 04:00~UT. 

Again we observe ripples with similar characteristics as those observed for rotation~\#1 (variations in the total magnetic field, with correlated or anti-correlated profiles in the vector components, separated by abrupt changes in the field orientation and strong fluctuations in temperature and density). They are located in the region between the ME of CME2.4+5 and after the shock-sheath from CME2.4+5 and SBC (marked with red arrows in Figure~\ref{insitu}).

\begin{table*}
\tiny 
\centering
    \begin{tabular}{|l|l|l|l|l|l|l|l|l|}
    \hline
        Event no. & CME1.1 & CME1.2  & CME1.3  & CME1.4  & CME2.1 & CME2.2 & CME2.3 & CME2.4+5 \\  \hline
        Date & Nov-01 & Nov-02  & Nov-03  & Nov-03  & Nov-27 & Nov-28 & Nov-28 & Nov-29 \\  
        SR location$^i$ & S30E20 & N25E25  & S20W30   & N15W30  &  W30S30  & N20E10  & N25E40  &  S15W00 \\  
        Flare class &  -- &  -- & M1.7/C2.8  & C3.2  &  --  &  C6.7 & C5.6  & M3.4+M9.8 \\  
        FR chirality & right & left  &  -- & left  & right   &  left & left  &  right \\  
        Filament relation &  yes &  yes &  no & yes (huge)  &  yes & no  &  yes & no \\  
        CME PA in WL &  160--200 & 0--70 &  260--280 & partial halo &  150--250  & partial halo  &  0--90 & full halo \\ CME GCS speed [km/s] & 650$\pm$30 & 630$\pm$30 & 350$\pm$20 & 1190$\pm$60 & 550$\pm$30 & 520$\pm$30 & 1020$\pm$50 & 640$\pm$30 \\   
        \hline
        CH area [$\times$10$^{10}$ km$^2$] & 8.5 & 8.5 & 8.5 & 8.5 & 10.1 & 10.1 & 10.1 & 10.1 \\ 
        CH B-signed [G] & $-$5.2 & $-$5.2 & $-$5.2 & $-$5.2 & $-$3.5 & $-$3.5 & $-$3.5 & $-$3.5 \\ 
        CHIP (apex) [G] & 1.77 & 1.77 & 0.68 & 0.76 & 0.27$^+$ & 0.46 & 0.70 & 0.81 \\ \hline
        3D DBM hit/miss & miss  & clear miss  & likely miss  &  likely hit &  likely miss  &  likely hit & clear miss  & clear hit \\  
        3D DBM FR arrival [UT] & --  & --  & --  &  Nov-05 23:00 &  --  &  Dec-02 00:44 & --  & Dec-02 00:44 \\  
        3D DBM shock [km/s] & --  & --  & --  &  Nov-05 13:00 &  --  &  Dec-01 14:44 & --  & Dec-01 14:44 \\ 
        \hline
        Shock arrival [UT] & --  & --  & Nov-05 09:15  &  Nov-05 12:30 &  --  &  Dec-01 00:22 & --  & Dec-01 09:38 \\  
        FR arrival [UT] & -- & -- & -- & -- & -- & Dec-01 07:28 & -- & Dec-01 20:45 \\  
        FR speed [km/s] & -- & -- & -- & -- & -- & 460 & -- & 430 \\  
        FR chirality & --  & --  & --  &  -- &  --  &  -- & --  & right \\ \hline
    \end{tabular}
    \caption{Summary Table. The date is for the year 2023. Block 1: CME source region (SR) characteristics with $^i$: to the nearest 5° at the launch time of CME (FR: flux rope; PA in WL: projected position angle of the CME as seen in white-light LASCO data). 3D speed and error for each CME as derived from GCS reconstructions (rounded; see more details in Table~\ref{tiny_table}). Block 2: Derived CATCH parameters related to the CH and CHIP results (the CH area given is corrected for projection effects; B-signed refers to the signed mean magnetic field strength; +: the calculated value corresponds to instances when the source region is visible during CH CATCH time, yet part of the CME apex lies on the opposite side of the Sun due to the longitudinal separation between the SR and the CH). Block 3: Hit/miss likeliness of each CME derived from CME cone projections with GCS parameters and errors. 3D DBM results for the FR and shock (see Table~\ref{tiny_table} for details of 3D DBM input parameters). Block 4: In-situ measured FR and shock arrival times, and characteristics of the observed magnetic structure. } \label{summary_table}
\end{table*}

\subsection{The role of the heliospheric current sheet}

   \begin{figure*}
   \centering
   \includegraphics[width=0.85\textwidth]{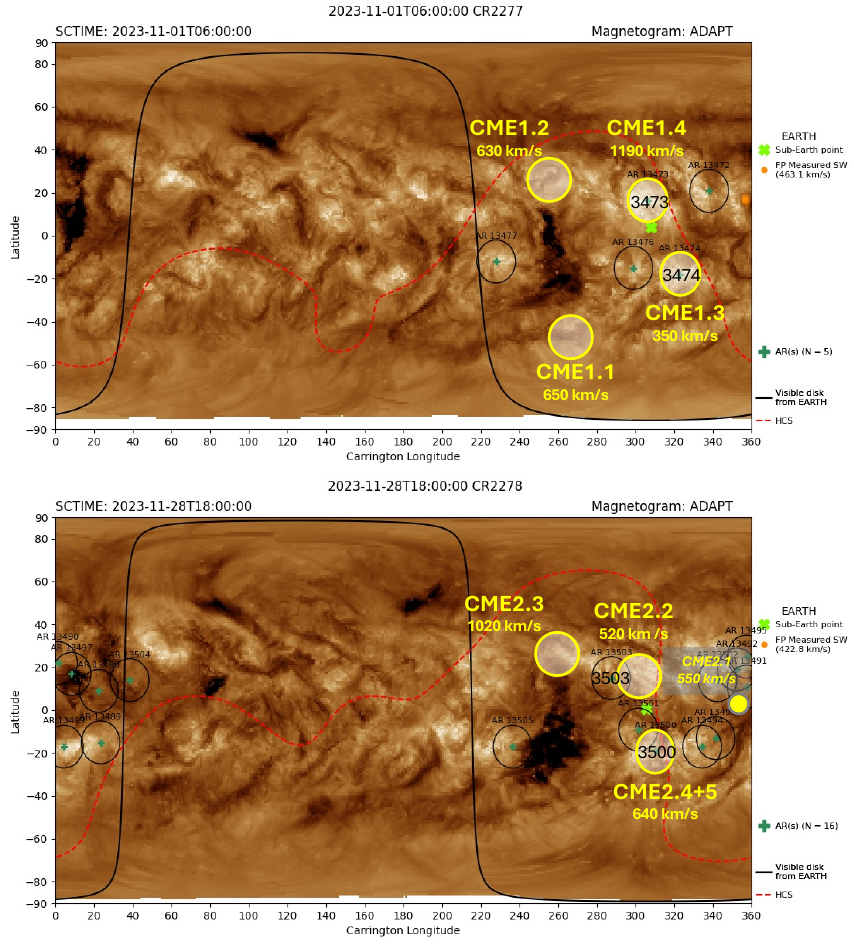}
      \caption{Location of the HCS and respective CME source regions (yellow circles) overlaid on SDO/AIA~193\AA~synoptic images. The HCS location is derived by the Magnetic Connectivity Tool on the basis of a PFSS \citep{Riley2006} extrapolation for a surface height of 3Rs. The results are provided by IRAP under \texttt{http://connect-tool.irap.omp.eu} \citep[see][]{Rouillard2020}.
    }
         \label{HCS_location}
   \end{figure*}
 
For relating the complex in-situ measurements with the observed SBCs back to the solar surface, we re-investigate the source regions on the Sun focusing on the HCS. The HCS is found to be rather inert in its reaction to activity \citep[e.g.,][]{hoeksema83,smith2001}. Potential field source surface \citep[PFSS;][]{Riley2006} extrapolations cover relatively well the large-scale structure of the interplanetary magnetic field \citep[see e.g.,][]{Owens2022}. In that respect, the HCS tilt close to the Sun should be described well by the extrapolation from the photospheric magnetic field. We derive the HCS location using the Magnetic Connectivity Tool\footnote{\url{http://connect-tool.irap.omp.eu}} \citep[see][]{Rouillard2020}, which is based on a PFSS extrapolation from ADAPT \citep[Air Force Data Assimilative Photospheric Flux Transport;][]{arge00} magnetograms.  The HCS is extracted at a surface height of 3\,Rs. The results are shown in Figure~\ref{HCS_location} with the derived HCS location as well as source regions overlaid on AIA~193\AA~images. We observe for CME1.3, CME1.4 and CME2.2 that their source regions are next to the HCS, which is highly tilted (North-South) in that regions. The very elongated filament channel related to CME1.4 crosses the HCS, meaning that the erupting filament and CME propagation/expansion behavior might be strongly influenced by it. In a similar way, CME2.2 revealing activity signatures over a large latitudinal region, was most likely influenced by the nearby HCS. Hence, for both episodes the location of the HCS relative to the solar source region of a CME seems to play an important role. As the HCS serves as magnetic obstacle the interaction with a CME might lead to strong compressional effects for the magnetic field structures involved. It is pointed out by \cite{Echer2004} that all CME magnetic structures and their shocks which are located at sector boundaries cause geomagnetic effects. 

\section{Discussion and Conclusions}\label{conclusion}

We analyze two comparable episodes covering analogous solar eruptive events and their geomagnetic impacts November 4--5, and December 1--2, 2023. The two episodes of events occurred in similar (active) regions on the Sun located near a CH. The study highlights the complexity of the heliospheric conditions during periods of interacting large-scale magnetic field structures that may lead to significant differences in geomagnetic effects. 

In total eight CMEs are investigated, 2 $\times$ 4 events, with the two episodes being separated by a complete solar rotation. Using stereoscopic white-light data, we reconstruct the 3D geometry, speed, and direction of motion for each CME. In addition, we derive the influence (such as deflection) of the nearby CH on each CME using the CHIP parameter \citep{Gopal2009}. The results serve as input for the CME propagation model 3D DBM (Dumbovic et al., 2024; to be submitted). This allows us to assess the impact probability of each event and to estimate their arrival times. With this we link between remote sensing image information and in-situ measurements and relate them to the observed geomagnetic effects.

Relating 3D DBM model results to in-situ measurements, we derive for rotation~\#1 a partial hit of CME1.4, while CME1.1 and CME1.2 would miss Earth and CME1.3 likely misses Earth. Indeed, the in-situ measurements reveal no obvious flux rope signature for the expected arrival time of CME1.4, but a short-lived magnetic structure presumably related to CME1.4. Two shocks are observed. While the second shock is linked to CME1.4, we may speculate that the first shock might be associated with CME1.3, although this is not fully supported by the results from the 3D DBM. However, a very recent study by \cite{Gil2024} focusing on the cosmic-ray peculiarities (ACRE event) observed November 5, 2023, would come to similar conclusions. For rotation~\#2 we derive from 3D DBM that CME2.2 and CME2.4+5 will both hit Earth with a likely interaction between the two CMEs. From the in-situ measurements two separate flux rope signatures are identified, with the most clear one connected to CME2.4+5. The shock of the fast CME2.4+5 might be located inside the ME of CME2.2 ahead. 

Relating the CME signatures on the Sun to the in-situ measurements, we observe for rotation~\#1 that the very energetic event CME1.4 corresponds to a massive filament eruption. The filament channel is found to cross the HCS which is highly tilted, i.e., oriented in the North-South direction. Also the source region of CME1.3 is found to be close to the HCS. Two shocks, the first one presumably from CME1.3 and the second from CME1.4 are located ahead and behind the sector boundary. For rotation~\#2, we obtain the source region of CME2.2 to be nearby the high-tilted HCS. Moreover, for rotation~\#2 the SBC is embedded within the ME from CME2.2, which is close in time to the shock component of CME2.4+5 that propagated into the ME of CME2.2. A statistical study on the geomagnetic effects of shocks inside the magnetic structures of CMEs is given in \cite{Lugaz2015}. Although CME1.4 only partially hit Earth, the observed stronger geomagnetic effects for rotation~\#1 in comparison to rotation~\#2 might stem from the interaction of several large-scale solar wind structures: shock-sheath components of CME1.3 and CME1.4 having strong negative B$_z$ values combined with the SBC and related structures \citep[see also][]{Echer2004, Crooker2004}. On the one hand, for both episodes the HCS serves as magnetic obstacle most likely affecting CME expansion and propagation behavior. On the other hand, CMEs may play an important role in the evolution of the HCS \citep[see recent PSP results by][]{romeo2023}. Hence, a local reconfiguration of the HCS in interplanetary space might indeed be observed at the location where the CME flank (CME1.4) is interacting with it \citep[see also][]{Gil2024}.

In the geomagnetic indices, we observe a three-step drop/increase in Dst/Hp30 for rotation~\#1 and a two-step drop/increase for rotation~\#2. Focusing on the Dst index, we find an absolute difference between the Dst minima from rotation~\#1 and rotation~\#2 of 55\,nT. This value roughly corresponds in rotation~\#1 to the first Dst drop of $-$54\,nT. That first drop in Dst is not related to any of the CMEs, but to an incoming high-density region with a rather strong negative B$_z$ component \citep[see also][]{Crooker2000}. Also for rotation~\#2 a high-density region following a strong negative B$_z$ component most likely contributed to the further drop in Dst. Individual HCS crossings have ``transition zones'' which might last longer than 24 hours and are related to specific magnetic structures often containing high plasma-$\beta$ regions and density blobs released from the top of the helmet streamer \citep[see e.g.,][]{Lepping1996, Sheeley1997, Lavraud2020}. 

Interestingly, after each SBC we find some distinct variations (``ripples'') in the magnetic field regions, which are characterized as short-term structures (mesoscale range of several hours) in the total magnetic field separated by abrupt changes in the field orientation. The variations in the magnetic field are related to strongly fluctuating temperature and density profiles. Such signatures could be related to either i) small-scale high-density magnetic field structures due to CME-solar wind interaction \citep{cappello24}, ii) shock wave - magnetic field interaction, hence, compression regions \citep[cf., e.g.,][]{Pitna2021}, iii) magnetic reconnection between the CME and the HCS changing the magnetic field topology of the flux rope \citep[e.g.,][]{Winslow2016}, or iv) periodic density structures in the mesoscale range as part of the slow solar wind \citep{Viall2021}.

Concluding, the HCS together with the extended HPS, and related sector boundary crossings play an important role in the geomagnetic effects of Earth-directed CMEs. But it is not only deflection or rotation that affects the CME when propagating near the HCS \citep{Yurchyshyn2001, Vourlidas2011, Isavnin2014, Kay2015}. For low-tilted (East-West) HCS, the ``same-opposite side effect" has influence on the impact strength of CMEs \citep{Henning1985}. A lowered geoeffectiveness of CMEs is observed for events launched on the opposite side of the HCS compared to Earth \citep{Zhao2007, Dumbovic2021}. However, for a high-tilted (North-South) HCS, CMEs which are launched close to it, may have an enhanced impact due to additional compression and deflection of the involved magnetic field structures. The orientation of the HCS is currently an under-represented parameter in studies relating solar activity phenomena to geomagnetic effects and might also need more attention in the applied research of Space Weather.

\begin{acknowledgements}
     We thank the anonymous reviewer for the instructive comments, which helped to improve the manuscript. We gratefully acknowledge the support from the Austrian-Croatian Bilateral Scientific Projects ``Multi-Wavelength Analysis of Solar Rotation Profile'' and ``Analysis of solar eruptive phenomena from cradle to grave'' under the project number OEAD HR 01/2024. M.T. and G.M.C. acknowledge support from the Young Researchers Program (YRP), project number AVO165300016. M.D., K.M. and  A.K.R. acknowledge support by the Croatian Science Foundation under the project IP-2020-02-9893 (ICOHOSS). K.M., A.K.R. and F.M. acknowledge support by Croatian Science Foundation in the scope of Young Researches Career Development Project Training New Doctoral Students. F.K. and M.T. acknowledge support from the Austrian Science Fund (FWF): P 33285-N. We thank the geomagnetic observatories (Kakioka [JMA], Honolulu and San Juan [USGS], Hermanus [RSA], Alibag [IIG]), NiCT, INTERMAGNET, and many others for their cooperation to make the provisional Dst index available. This work utilizes data produced collaboratively between AFRL/ADAPT and NSO/NISP. Solar Orbiter is a space mission of international collaboration between ESA and NASA, operated by ESA. Metis was built and operated with funding from the Italian Space Agency (ASI), under contracts to the National Institute of Astrophysics (INAF) and industrial partners. Metis was built with hardware contributions from Germany (Bundesministerium für Wirtschaft und Energie through DLR), from the Czech Republic (PRODEX) and from ESA.
\end{acknowledgements}

%
%
\appendix

\section{Supplementary material}\label{appendix}

\subsection{Additional GCS reconstructions}

\begin{figure*}
   \centering
   \includegraphics[width=0.88\textwidth]{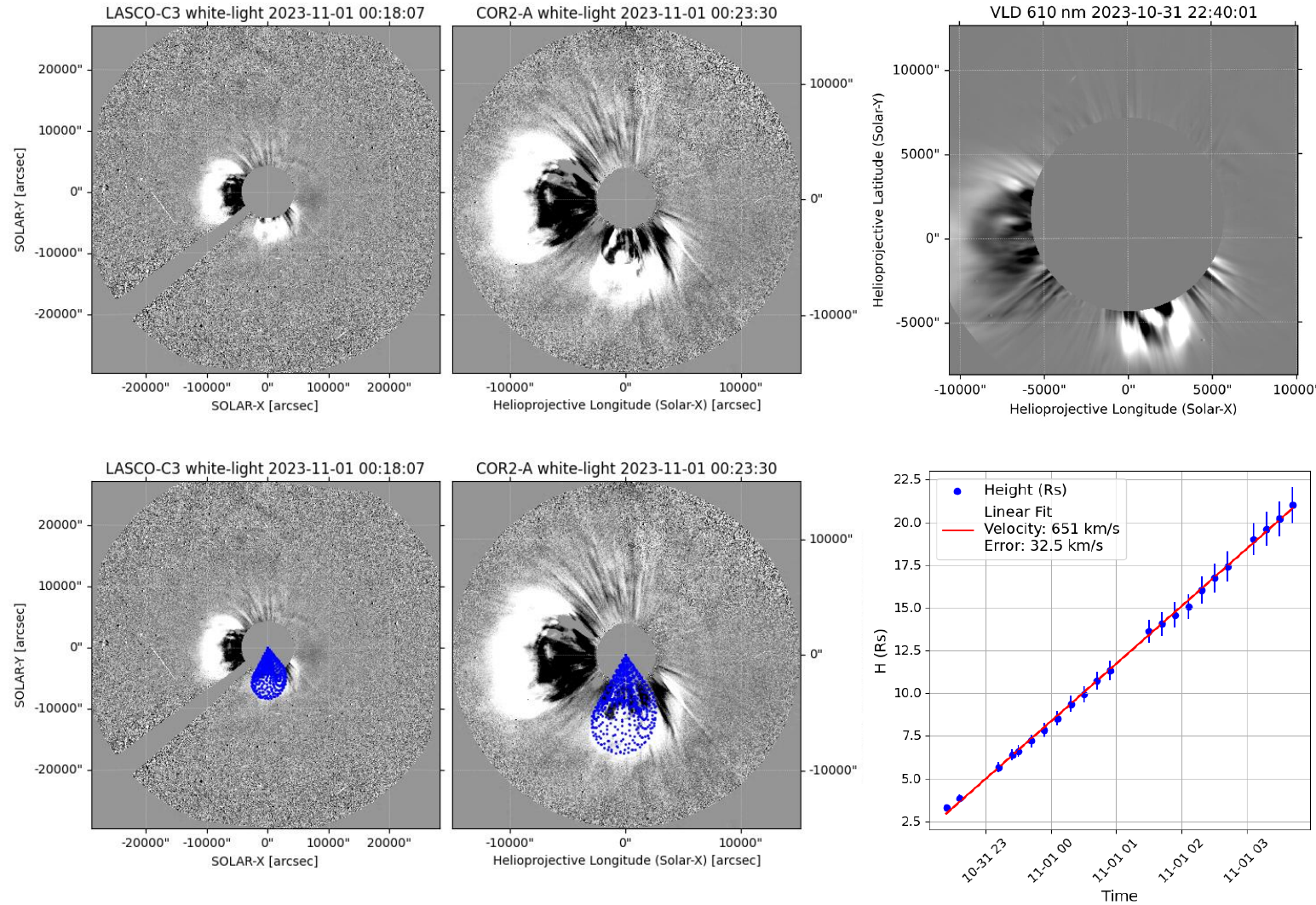}
      \caption{CME1.1: Multi-perspective white-light CME signatures (LASCO C2/C3, COR2, Metis), GCS reconstruction shown for a 3D height of 10.8Rs, and kinematics. From the height-time plot the speed is derived using a linear fit (given in the legend). 
}
         \label{CME1.1}
   \end{figure*}

\begin{figure*}
   \centering
   \includegraphics[width= 0.88\textwidth]{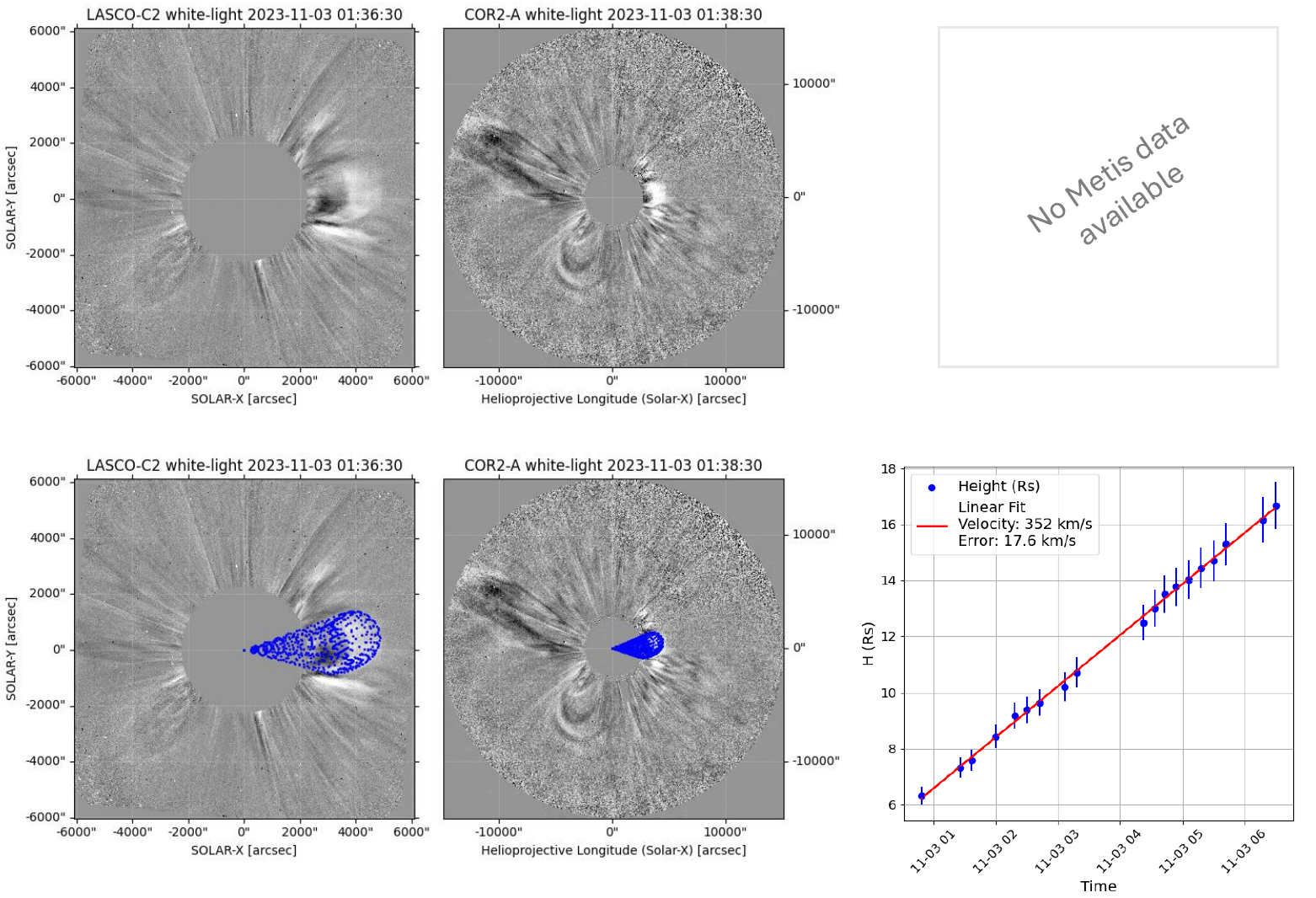}
      \caption{Same as Figure~\ref{CME1.1} but for CME1.3. GCS reconstruction is shown for a 3D height of 6.5Rs. 
}
         \label{CME1.3}
   \end{figure*}

\begin{figure*}
   \centering
   \includegraphics[width= 0.88\textwidth]{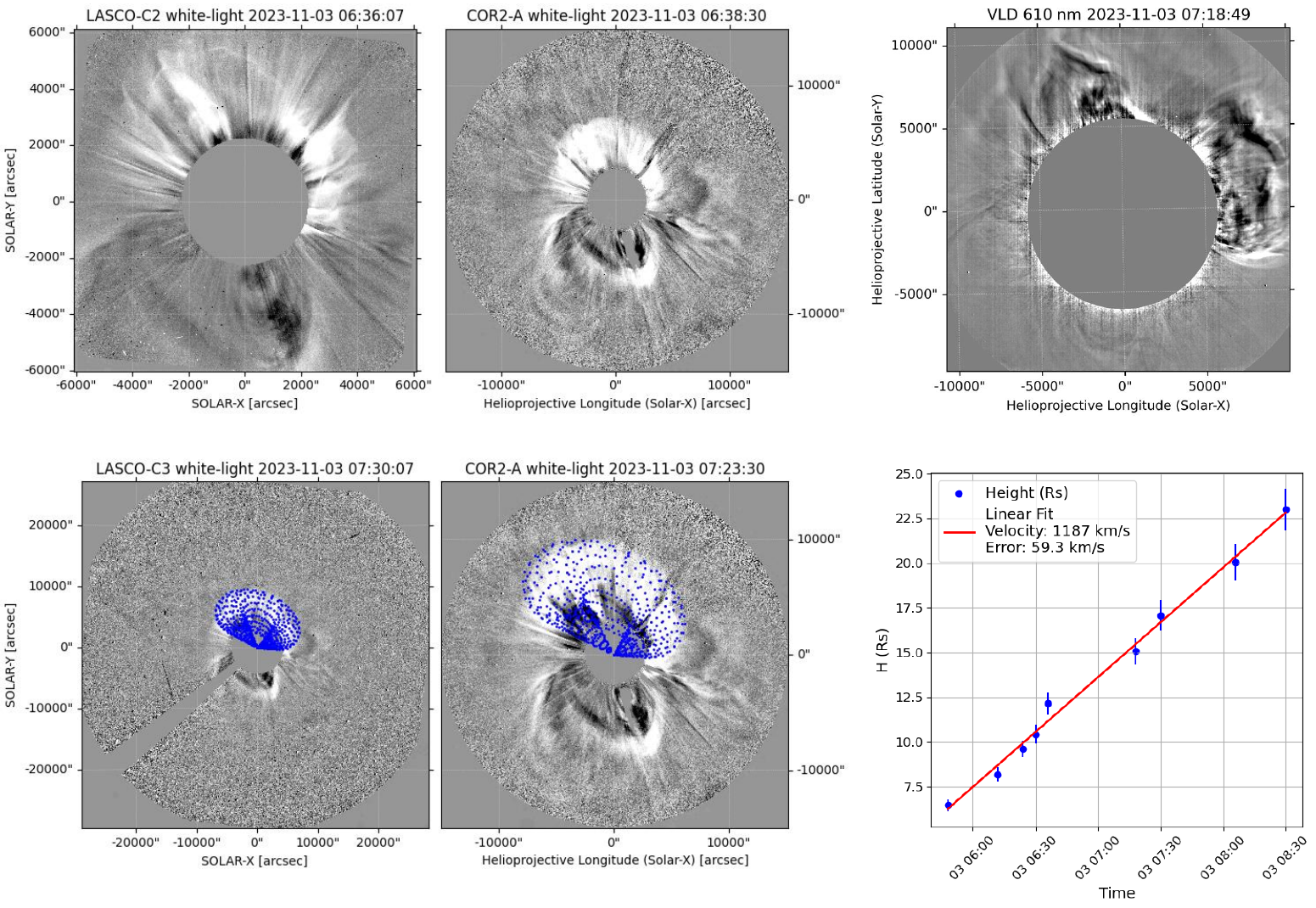}
      \caption{ Same as Figure~\ref{CME1.1} but for CME1.4. GCS reconstruction is shown for a 3D height of 9.4Rs.
}
         \label{CME1.4}
   \end{figure*}

\begin{figure*}
   \centering
   \includegraphics[width= 0.88\textwidth]{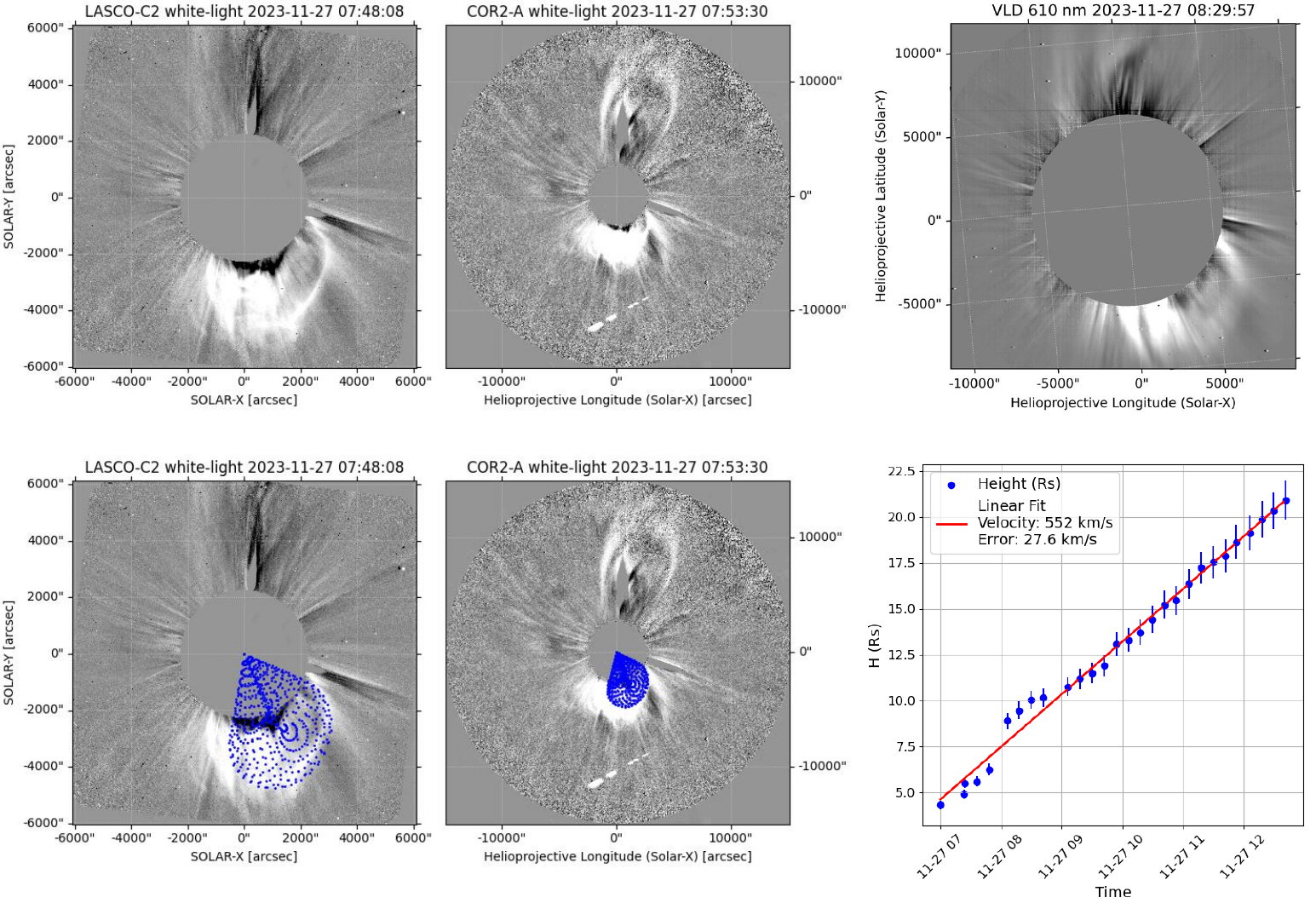}
      \caption{ Same as Figure~\ref{CME1.1} but for CME2.1. GCS reconstruction is shown for a 3D height of 13.8Rs.
}
         \label{CME2.1}
   \end{figure*}

\begin{figure*}
   \centering
   \includegraphics[width= 0.88\textwidth]{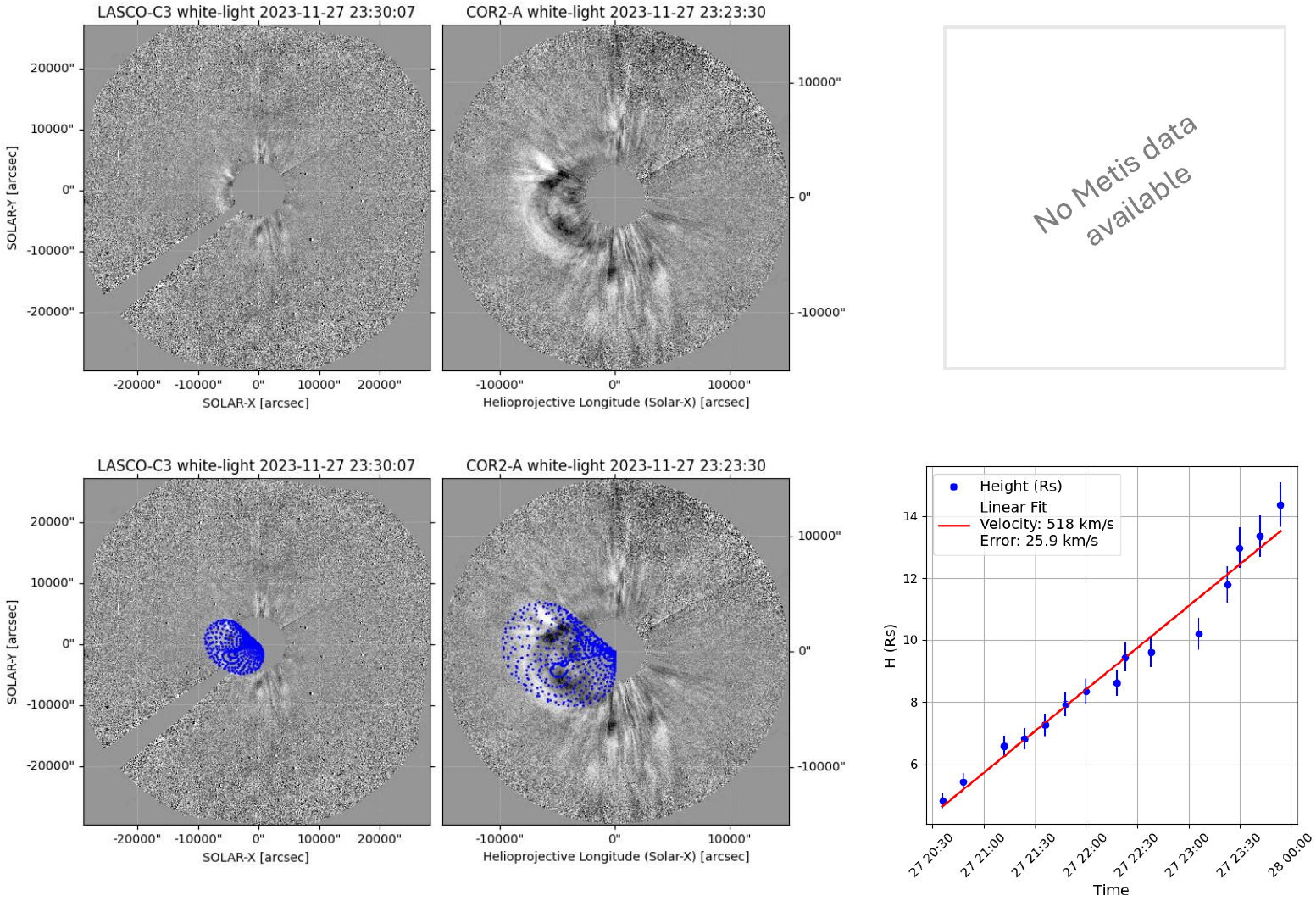}
      \caption{  Same as Figure~\ref{CME1.1} but for CME2.2. GCS reconstruction is shown for a 3D height of 6.6Rs. We note that we further constrain the GCS reconstruction by taking the low coronal signatures (see Figure~\ref{over2}) into account. 
}
         \label{CME2.2}
   \end{figure*}

\begin{figure*}
   \centering
   \includegraphics[width= 0.88\textwidth]{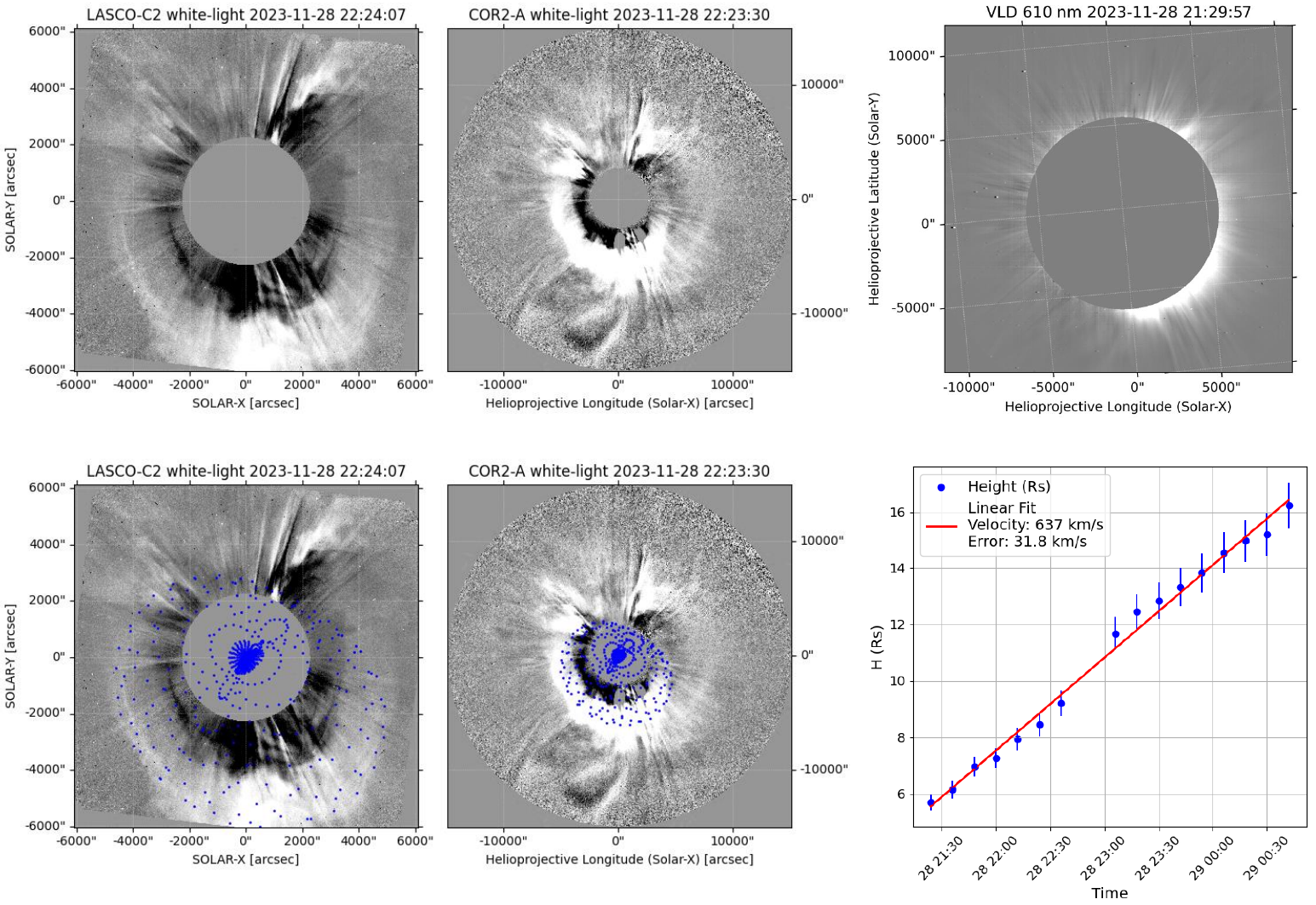}
      \caption{  Same as Figure~\ref{CME1.1} but for CME2.4+5. GCS reconstruction is shown for a 3D height of 13.6Rs. We note that due to the almost spherical appearance we cannot give a strong constraint on the tilt.
}
         \label{CME2.4+5}
   \end{figure*}

Figures~\ref{CME1.1}--\ref{CME2.4+5} show the GCS reconstructions for CME events from rotation~\#1 and~\#2. The GCS is outlined on simultaneously matching the outer front of the CME flux rope part as observed in STEREO-A, SOHO, and Solar Orbiter (Metis VLD610; whenever available) white-light images at roughly the same time. For each event several time steps were chosen to perform GCS from which a height-time plot is given. Using a linear fit, the speed of the CME is derived with error estimates taken from the statistical study by \cite{Verbeke2023}.


\subsection{In-situ measurements - zoom-in}
\begin{sidewaysfigure*}
   \centering
   \includegraphics[width=20cm]{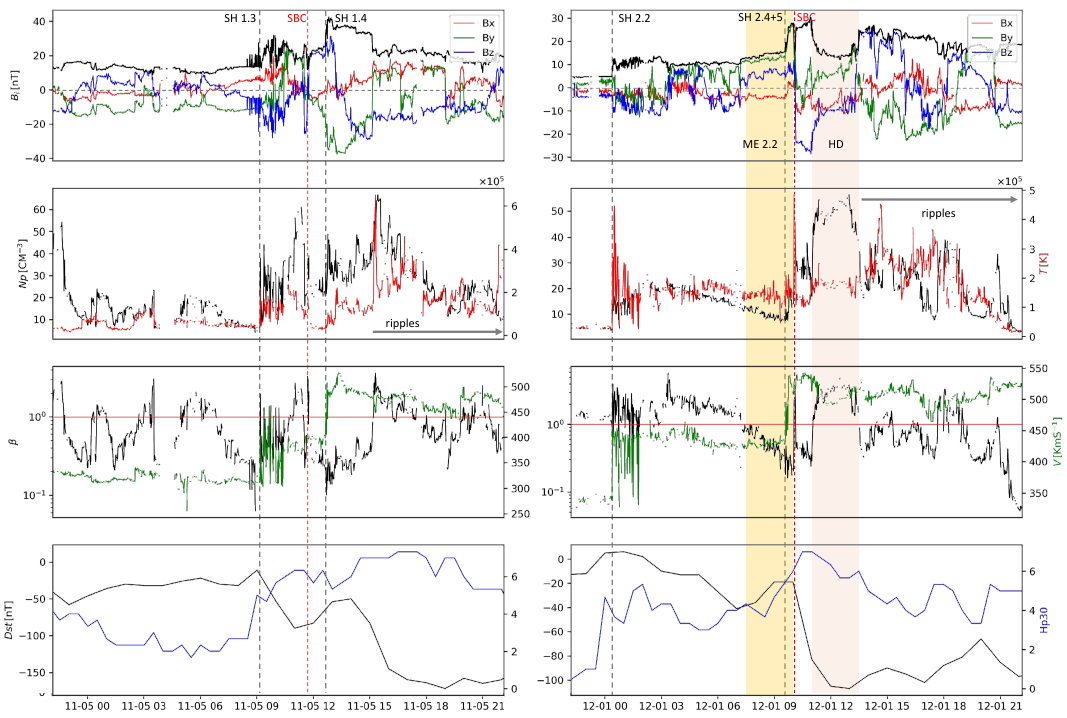}
      \caption{Zoom-in of Figure~\ref{insitu}. See description and legend in Figure~\ref{insitu}.
}
         \label{zoom-insitu}
   \end{sidewaysfigure*}

Figure~\ref{zoom-insitu} shows in detail the region around the SBC. The SBC is identified by a strong drop in the total magnetic field, together with a change in the polarity for the B$_x$ (positive to negative) and B$_y$ (positive to negative) component. Plasma-$\beta$ and density show a clear short-duration peak (spike) at the same time. These criteria hold for both rotations (rotation~\#1 - left panel; rotation~\#2 - right panel). In addition we indicate the start of the ripples after. Labels given in Figure~\ref{zoom-insitu} are the same as for Figure~\ref{insitu}.

\subsection{Detailed Summary Table}\label{appendix_sum}
Table~\ref{tiny_table} gives a summary of the GCS parameters derived for each CME (see Figures~\ref{CME1.2} and \ref{CME2.3}, and Appendix Figures \ref{CME1.1} to \ref{CME2.4+5}). From the GCS results we identified events which are most likely related to in-situ signatures, i.e., CME1.4, CME2.2 and CME2.4+5. For these CMEs we performed 3D DBM runs to assess arrival time and speed for better linking and interpretation of the in-situ signatures. As input for the 3D DBM runs we use the results from GCS for CME2.2 and CME2.4+5 and tweaked the results from GCS for CME1.4 such to come from a glancing blow to a flank hit. The optimum values for the interaction events produces an arrival which is roughly 10 hours later than observed. We are able to find a set of parameters within the GCS errors that produce the observed signatures in-situ.

\begin{table*}[]
\tiny
\begin{tabular}{l|r|r|r|r|r|r|r|r|}
GCS params.          & \multicolumn{1}{l}{CME1.1}           & \multicolumn{1}{l}{CME1.2}     & \multicolumn{1}{l}{CME1.3}      & \multicolumn{1}{l}{CME1.4}                 & \multicolumn{1}{l}{CME2.1}        & \multicolumn{1}{l}{CME2.2}           & \multicolumn{1}{l}{CME2.3}     & \multicolumn{1}{l}{CME2.4+5}         \\
\multicolumn{9}{c}{Results from GCS}                                                                                                                                                                                                                                                                                                          \\
Half angle {[}°{]}                & 14             & 15           & 5             & 30                                  & 15                                & 20                                   & 15                             & 44                                   \\
$k$ {[}rad{]}                       & 0.35                                 & 0.25                           & 0.14                            & 0.24                                       & 0.25                              & 0.35                                 & 0.31                           & 0.94                                 \\
latitude {[}°{]}                  & $-$55                                  & 50                             & 6                               & 29                                         & $-$40                               & $-$1                                   & 43                             & $-$23                                  \\
longitude {[}°{]}                 & 3                                    & 322                            & 35                              & 0                                          & 23                                & 330                                  & 312                            & 3                                    \\
tilt {[}°{]}                      & 60                                   & 40                             & 25                              & $-$20                                        & 60                                & $-$40                                  & $-$61                            & $-$35                                  \\
Time@21.5 Rs                   & Nov-01 03:50                  & Nov-02 09:42               & Nov-03 09:09                & Nov-03 08:00                           & Nov-27 12:54                  & Nov-28 02:33                     & Nov-28 03:14               & Nov-29 02:18                     \\
speed  {[}km/s{]}                 & 651$\pm$32                                  & 629$\pm$31                            & 352$\pm$18                            & 1187$\pm$59                                       & 552$\pm$28                               & 518$\pm$26                                  & 1020$\pm$51                           & 637$\pm$32                                  \\
hit/miss                          & \multicolumn{1}{l}{miss}             & \multicolumn{1}{l}{miss}       & \multicolumn{1}{l}{miss}        & \multicolumn{1}{l}{gla.-blow} & \multicolumn{1}{l}{miss}          & \multicolumn{1}{l}{hit}              & \multicolumn{1}{l}{miss}       & \multicolumn{1}{l}{hit}              \\ 
\multicolumn{9}{c}{Errors for GCS parameters}                                                                                                                                                                                                                                                                                               \\
Half angle {[}°{]}                & 10                                   & 10                             & 10                              & 10                                         & 10                                & 10                                   & 10                             & 10                                   \\
$k$ {[}rad{]}                       & 0.1                                  & 0.1                            & 0.1                             & 0.1                                        & 0.1                               & 0.1                                  & 0.1                            & 0.1                                  \\
latitude {[}°{]}                  & 6                                    & 6                              & 6                               & 6                                          & 6                                 & 6                                    & 6                              & 6                                    \\
longitude {[}°{]}                 & 11                                   & 11                             & 11                              & 11                                         & 11                                & 11                                   & 11                             & 11                                   \\
tilt {[}°{]}                      & 25                                   & 25                             & 25                              & 25                                         & 25                                & 25                                   & 25                             & 25                                   \\
\multicolumn{9}{c}{GCS parameters used for 3D DBM to simulate a best hit scenario within errors}                                                                                                                                                                                                                                                                    \\
Half angle {[}°{]}                & 24                                   & 25                             & 15                              & 40                                         & 25                                & 30                                   & 25                             & 54                                   \\
$k$ {[}rad{]}                       & 0,45                                 & 0,35                           & 0,24                            & 0.34                                       & 0.35                              & 0.45                                 & 0.41                           & 0.99*                                 \\
latitude {[}°{]}                  & $-$49                                  & 44                             & 0                               & 23                                         & $-$34                               & $-$1*                                   & 37                             & $-$17                                  \\
longitude {[}°{]}                 & 3*                                    & 333                            & 24                              & 0*                                          & 12                                & 341                                  & 323                            & 3*                                    \\
tilt {[}°{]}                      & 85                                   & 15                             & 0                               & $-$45                                        & 85                                & $-$15                                  & $-$61                            & $-$60                                  \\
hit/miss                          & \multicolumn{1}{l}{gla.-blow}    & \multicolumn{1}{l}{miss}       & \multicolumn{1}{l}{flank hit}   & \multicolumn{1}{l}{full hit}               & \multicolumn{1}{l}{flank hit}     & \multicolumn{1}{l}{hit}              & \multicolumn{1}{l}{miss}       & \multicolumn{1}{l}{hit}              \\
\multicolumn{9}{c}{GCS parameters used for 3D DBM to simulate a worst hit scenario within errors}                                                                                                                                                                                                                                                                   \\
Half angle {[}°{]}                & 4                                    & 5                              & 5*                               & 20                                         & 5                                 & 10                                   & 5                              & 34                                   \\
$k$ {[}rad{]}                       & 0.25                                 & 0.15                           & 0.1*                             & 0.14                                       & 0.15                              & 0.25                                 & 0.21                           & 0.84                                 \\
latitude {[}°{]}                  & -61                                  & 56                             & 12                              & 35                                         & $-$46                               & $-$7                                   & 49                             & $-$29                                  \\
longitude {[}°{]}                 & 14                                   & 311                            & 46                              & 11                                         & 34                                & 319                                  & 301                            & 14                                   \\
tilt {[}°{]}                      & 35                                   & 65                             & 50                              & 5                                          & 35                                & $-$65                                  & $-$86                            & $-$10                                  \\
hit/miss                          & \multicolumn{1}{l}{miss}             & \multicolumn{1}{l}{miss}       & \multicolumn{1}{l}{miss}        & \multicolumn{1}{l}{miss}                   & \multicolumn{1}{l}{miss}          & \multicolumn{1}{l}{miss}             & \multicolumn{1}{l}{miss}       & \multicolumn{1}{l}{hit}              \\ \hline
Conclusion:                           & \multicolumn{1}{l}{miss} & \multicolumn{1}{l}{clear miss} & \multicolumn{1}{l}{likely miss} & \multicolumn{1}{l}{likely hit}             & \multicolumn{1}{l}{miss}   & \multicolumn{1}{l}{likely hit}       & \multicolumn{1}{l}{clear miss} & \multicolumn{1}{l}{clear hit}        \\ \hline
\multicolumn{9}{c}{3D DBM runs: input and results for arrival time and speed at 1AU}                                                                                                                                                                                                                                                        \\  
speed  {[}km/s{]}                 & N/A                                  & N/A                            & N/A                             & 1187                                       & N/A                               & 518              &  N/A           & 637              \\
radius@21.5Rs                   & N/A                                  & N/A                            & N/A                             & 4.1                                        & N/A                               & 5.4              & N/A           & 9.8              \\
density ratio                     & N/A                                  & N/A                            & N/A                             & 4                                          & N/A                               & 2                & N/A         & 2               \\
$\gamma$ {[}1e-7km-1{]}               & N/A                                  & N/A                            & N/A                             & 0.25                                       & N/A                               & 0.34             & N/A          & 0.19             \\
$w$ {[}km/s{]}                      & N/A                                  & N/A                            & N/A                             & 350                                        & N/A                               & 350              & N/A          & 430            \\
FR TT {[}h{]}                 & N/A                                  & N/A                            & N/A                             & 63                                         & N/A                               & \multicolumn{3}{c}{94.2}                                                                                     \\
FR v {[}km/s{]}               & N/A                                  & N/A                            & N/A                             & 485                                        & N/A                               & \multicolumn{3}{c}{415}                                                                                      \\
FR arrival                    & N/A                                  & N/A                            & N/A                             & Nov-05 23:00                           & N/A                               & \multicolumn{3}{c}{Dec-02 00:44}                                                                         \\
shock           & N/A                                  & N/A                            & N/A                             & Nov-05 13:00                           & N/A                               & \multicolumn{3}{c}{Dec-01 14:44}                                                                         \\ 

\end{tabular}
\caption{CME hit/miss statistics with GCS parameters derived for each CME and 3D DBM runs performed for those CMEs that likely hit Earth. The date is for the year 2023. Values marked with * are those where the full error range could not be applied, as e.g., it would mean that the aspect ratio is larger than 1 which is physically not meaningful, or that lon/lat goes too far away from the apex. The 3D DBM runs were performed using unaltered input from GCS results for rotation\#2, but for CME1.4 adapted within the derived GCS errors to generate a flank hit (lat=25, tilt=$-$35) from which arrival time and speed could be obtained.}\label{tiny_table}
\end{table*}

\section{3D DBM modeling of the CME-CME interaction}\label{appendix2}

3D DBM tracks the kinematics of the infinitesimally thin leading edge of the CME. The interaction of two CMEs in 3D DBM can therefore be approximated as a trailing front encountering the leading front. We note that in reality this is not the case, because the front of the trailing CME first interacts with the back of the leading CME. Nevertheless, this approach offers a simple and fast way to treat CME-CME interactions with 3D DBM. We consider the point where the kinematic curves of two CMEs cross as the interaction point. We assume that during the interaction impulse and mass are conserved and that after the interaction the two CMEs move on as a single entity. The initial parameters are then recalculated for the entity at the interaction point (starting time, starting distance, speed based on conserved momentum, $\gamma$ parameter), with the solar wind speed slightly enhanced. The $\gamma$ parameter is recalculated based on the increased CME leading front cross sectional area and mass \citep[for details see][]{Guo2018,Dumbovic2019}:
\begin{equation}
\gamma_{1+2} = \frac{\gamma_1+\frac{M_1}{M_2}\mathrm{sin}^2\omega_{1+2}}{(\frac{M_1}{M_2}+1)\mathrm{sin}^2\omega_1},
\label{eq-gamma_entity}
\end{equation}

\noindent where $\gamma_1$ is the original $\gamma$ parameter of the first CME, $\frac{M_1}{M_2}$ is the mass ratio of the two CMEs (estimated roughly by the observer, based on the impulsiveness, brightness and spatial extent of the two CMEs in the white light observations), $\omega_1$ is the spatial extent of the first CME (estimated as the extent in either equatorial or meridional plane, whichever is larger) and $\omega_{1+2}$ is the extent of the merged CME entity, which is derived by combining the extents of both CMEs in equatorial and meridional planes and choosing whichever is larger. The extent (half-width) of the entity in equatorial plane is calculated as:
\begin{equation}
	\omega_{1+2} =
	\begin{cases}
    \frac{1}{2}(\omega_{\mathrm{max}}+lon_{\mathrm{max}}-lon_{\mathrm{min}}+\omega_{\mathrm{min}}),\\
    \mathrm{when} \,\,lon_{\mathrm{max}}-\omega_{\mathrm{max}} >lon_{\mathrm{min}}-\omega_{\mathrm{min}}\\
    \omega_{\mathrm{max}}, \mathrm{otherwise}
	\end{cases}
\label{eq-extent}
\end{equation}

\noindent where $\omega$ stands for half-width, $lon$ for longitude and indices $max$ and $min$ correspond to wider and narrower CME in the interaction, respectively. An analogous equation is used to calculate the half-width of the entity in the meridional plane.

The original drag parameters corresponding to CMEs before interaction are calculated according to \cite{cargill04} and \cite{Vrsnak2014} as:
\begin{equation}
\gamma=\frac{2}{\pi}\frac{1}{(\frac{n}{n_\mathrm{SW}}+\frac{1}{2})}\frac{C_\mathrm{d}}{r}
\,,
\label{eq-gamma_cylinder}
\end{equation}

\noindent where we assume $C_d$ equals 1, $r$ is the radius of the CME, which has toroidal geometry and is calculated according to \citet{thernisien11}. The density ratio $n/n_\mathrm{SW}$ is estimated based on the CME speed, using the CME mass vs.\,speed relation \citep[see Table 1 in ][]{dissauer19} and typical values for density ratio used in the ENLIL cone model \citep[see e.g.,][]{yordanova24}, grouping CMEs into three classes: slow ($<$750~km/s), intermediate (750--1500~km/s) and fast ($>$1500~km/s). For intermediate CMEs we use density ratio 4 (typical value used in ENLIL cone). For slow and fast CMEs with masses $\approx$ 2 times smaller and 1.5 times larger than for intermediate CMEs, respectively, we use density ratios of 2 and 6, respectively.



\bibliographystyle{aa.bst}

\end{document}